\documentclass[runningheads,fleqn]{cl2emult}

\usepackage{makeidx}  
\usepackage{graphicx} 
\usepackage{subeqnar} 
\usepackage{multicol} 
\usepackage{cropmark} 
\usepackage{physmu}   
\makeindex            
\begin{document}
\title*{Stability of Spatial Optical Solitons}
\author{Yuri S. Kivshar
\and Andrey A. Sukhorukov}
\authorrunning{Yuri S. Kivshar and Andrey A. Sukhorukov}

\maketitle              

\section{Introduction}

{\em Spatial optical solitons} are known to originate from the nonlinearity-induced diffraction suppression and beam self\--trapp\-ing in a bulk dielectric medium~\cite{general}. Since, generally speaking, the effect of diffraction is strong, considerable optical nonlinearities are required in order to compensate for the diffraction-induced beam spreading. As a result, in the vicinity of a self-trapped beam the refractive index experiences large deviations from a Kerr-type dependence and the models of generalized nonlinearities, that describe beam self-trapping phenomena and spatial solitons, become {\em nonintegrable}. 

For solitary waves of nonintegrable nonlinear models, {\em linear stability} is one of the crucial issues, since only stable (or weakly unstable) self-trapped beams can be observed in experiment. The study of soliton stability in nonlinear optical materials has a long history, also associated with the study of nonlinear waves in other media such as plasmas and fluids. Stability of one-parameter solitary waves has already been well understood for both fundamental (single-hump and nodeless) solitons~\cite{VK,Weinstein,ZK,makhan} and solitons with nodes and multiple humps~\cite{jones,Gr}. The pioneering results of N.~Vakhitov and A.~Kolokolov~\cite{VK}, known these days as {\em the Vakhitov-Kolokolov stability criterion}, found their rigorous justification in a general mathematical theory developed by M.~Grillakis {\em et al.}~\cite{Gr1}. Although the corresponding stability and instability theorems for the scalar nonlinear Schr\"odinger (NLS) models formally extend to the case of multi-parameter solitons~\cite{Gr1}, most of the examples analyzed so far correspond to solitary waves with {\em a single parameter}. Only very recently a systematic analysis of more general cases has been carried out, in connection with the study of three-wave parametric solitons in quadratic (or $\chi^{(2)}$) optical media~\cite{chi2-review}.

Recent progress in the study of soliton instabilities can be associated with the application of {\em a multi-scale asymptotic bifurcation theory} developed for {\em weakly unstable} stationary nonlinear localized waves. In the framework of this theory, the unstable eigenvalue of the associated linear spectral problem is treated as {\em a small parameter} of the asymptotic expansions~\cite{P1}, and the resulting equation for the
slowly varying soliton propagation constant may also account for weakly nonlinear effects that describe the long-time evolution of linearly unstable solitary waves, and an effective saturation of the soliton instability due to higher-order nonlinear effects. In the case of multi-parameter solitary waves, a simplified version of this asymptotic method applied near a marginal stability point (or, in general, a surface) is reduced to finding certain determinants constructed from the derivatives of the system invariants~\cite{three,P3,multi,skryabin}. However, the validity of this bifurcation theory has no rigorous proof, and it can only be used to {\em estimate} the domains of the soliton stability and instability. Since more general {\em oscillatory instabilities} may also occur~\cite{Gr,gap,coupled2,mih,johan}, numerical simulations are often required in order to verify the predictions of the asymptotic theory (see, e.g., Ref.~\cite{multi} as an example).

Recently, a general matrix criterion for the stability and instability of {\em multi-component solitary waves} was derived~\cite{rigorous} for a system of $N$ incoherently coupled NLS equations. In this general approach, the soliton stability is studied as a constrained variational problem reduced to finite-dimensional linear algebra. Unstable eigenvalues of the linear stability problem for multi-component solitary waves were shown to be connected with negative eigenvalues of the Hessian matrix constructed for the energetic surface of $N-$component spatially localized stationary solutions.

The analysis and, correspondingly, stability criteria obtained in Ref.~\cite{rigorous} can be extended, at least in principle,  to other types of solitary waves, such as incoherent solitons in non-Kerr media, parametric solitary waves in $\chi^{(2)}$ optical  media, etc. In all such cases, the results on stability and instability of solitons can be readily obtained with {\em a rigorous generalization} of some of the previously known results of the multi-scale asymptotic theory. However, in each of those cases some additional analysis is required, in order to clarify whether the results of the asymptotic theory completely define the stability properties of multi-component solitary waves. Beyond the validity of the multi-scale analysis, oscillatory instabilities may occur, and appropriate studies should rely solely on the numerical analysis of the corresponding eigenvalue problems.

In this Chapter, we present {\em a brief overview of the basic concepts} of the soliton stability theory and discuss some characteristic examples of the instability-induced soliton dynamics, in application to spatial optical solitons described by the NLS-type models and their generalizations.   First of all, we demonstrate a crucial role played in the stability theory by the soliton internal modes that present an important characteristic of solitary waves in nonintegrable nonlinear models. In particular, we study an example model of the NLS equation with higher-order nonlinearity, in order to show that the soliton internal mode gives birth to the soliton instability. Near the marginal stability point, the soliton stability can be analyzed by a multi-scale asymptotic technique when the instability growth rate is treated as a small perturbation parameter. We also discuss some results of the rigorous linear stability analysis of fundamental solitary waves and nonlinear impurity modes. More recent studies of higher-order solitons revealed that the so-called {\em multi-hump vector solitary waves} may become stable in some nonlinear models, and we discuss the stability of (1+1)-dimensional two- and three-hump composite solitons created by incoherent interaction of two optical beams in a photorefractive nonlinear medium. The final part of this Chapter is devoted to the stability of solitons in higher dimensions and, in particular, it discusses very recent results on the symmetry-breaking instability of a  (2+1)-dimensional vortex-mode composite soliton and the formation of a rotating dipole-like structure associated with a robust radially asymmetric {\em dipole-mode vector soliton}, a new composite object resembling a ``molecule of light''.

\section{Linear eigenvalue problem} \label{sec:linear_eigen}

To discuss the stability properties of spatial optical solitons, we consider the nonintegrable dimensionless generalized NLS equation that describes the (1+1)-dimensional beam self-focusing in a waveguide geometry, 
\begin{equation} \label{eq:NLS}
   i \frac{\partial \psi}{\partial z} 
   + \frac{\partial^2 \psi}{\partial x^2} 
   + {\cal F}(I; x) \psi 
   = 0 ,
\end{equation}
where $\psi(x,z)$ is the dimensionless complex envelope of the electric field, $x$ is the transverse spatial coordinate, $z$ is the propagation distance, $I = |\psi(x,z)|^2$ is the beam intensity, and the real function ${\cal F}(I; x)$ characterises both linear and nonlinear properties of a dielectric medium, for which we assume that ${\cal F}(0; \pm \infty ) = 0$. Stationary spatially localized solutions of the model~(\ref{eq:NLS}) have the standard form, $\psi(x,z) = \Phi(x; \beta) e^{i \beta z}$, where $\beta$ is the soliton propagation constant ($\beta>0$) and real function $\Phi(x; \beta)$ vanishes for $|x| \rightarrow \infty$.  An important conserved quantity of the soliton in the model~(\ref{eq:NLS}) is its {\em power} defined as 
\begin{equation} \label{power}
   P(\beta) 
   = \int_{-\infty}^{+\infty} |\psi(x,z)|^2 d x 
   = \int_{-\infty}^{+\infty} \Phi^2(x; \beta) d x .
\end{equation}

To find the linear stability conditions,  we consider the evolution of a small-amplitude perturbation of the soliton presenting the solution in the form
\begin{equation} \label{eq:psi}
 \psi (x,z) 
 = \left\{ \Phi(x; \beta) 
           + \left[ v(x) - w(x) \right] e^{i \lambda z} 
           + [v^{\ast}(x) + w^{\ast}(x)] e^{-i \lambda^{\ast} z} \right\}
              e^{i \beta z} ,
\end{equation}
where the star stands for a complex conjugation, and obtain the linear eigenvalue problem for $v(x)$ and $w(x)$,
\begin{equation} 
\label{eq:linear_eigen}
 \begin{array}{l} {\displaystyle
  L_0 w = \lambda v, \quad L_1 v = \lambda w ,
 } \\*[9pt] {\displaystyle
  L_j = - \frac{d^2}{d x^2} + \beta - U_j , 
 } \end{array}
\end{equation}
where $U_0 = {\cal F}(I; x)$ and $U_1 = {\cal F}(I; x) + 2 I [\partial {\cal F}(I; x) / \partial I]$.

A stationary solution of the model~(\ref{eq:NLS}) is stable if all the eigenmodes of the corresponding linear problem~(\ref{eq:linear_eigen})
do not have exponentially growing amplitudes, i.e. ${\rm Im}(\lambda) = 0$. It can be demonstrated that the continuum part of the linear spectrum of the problem~(\ref{eq:linear_eigen}) consists of {\em two symmetric branches} corresponding to real eigenvalues with the absolute values $|\lambda| > \beta$, and therefore only discrete eigenstates are responsible for the stability properties of localized waves. 
Then, the corresponding eigenmode solutions fall into the following categories:
\begin{itemize}
\item {\em internal modes} with real eigenvalues describe periodic oscillations;
\item {\em instability modes} correspond to purely imaginary eigenvalues; 
\item {\em oscillatory instabilities} can occur when the eigenvalues are complex.
\end{itemize}

In what follows, we present several approaches that allow us to study analytically and numerically the structure of the discrete spectrum in order to determine the linear stability properties of solitary waves. We also analyze the nonlinear evolution of unstable solitons.

\section{Soliton internal modes and stability} \label{sec:imode}

Since the soliton instabilities always occur in nonintegrable models, it is interesting to know what kind of distinct features of the solitary waves in nonintegrable models might be responsible for their instabilities. It is commonly believed that solitary waves of nonintegrable nonlinear models differ from solitons of integrable models only in the character of the soliton interactions: unlike ``proper'' solitons, interaction of solitary waves is accompanied by radiation~\cite{kivmal}. However, the soliton instabilities are associated with nontrivial effects of different nature that are generic for localized waves of nearly integrable and nonintegrable models. In particular, a small perturbation to an integrable model may create {\em an internal mode} of a solitary wave~\cite{prl_internal}. This effect is beyond a regular perturbation theory, because solitons of integrable models do not possess internal modes. But in nonintegrable models such modes may introduce {\em qualitatively new features} into the system dynamics and, in particular, lead to the appearance of the soliton instabilities.

\begin{figure}
\includegraphics[width=.8\textwidth]{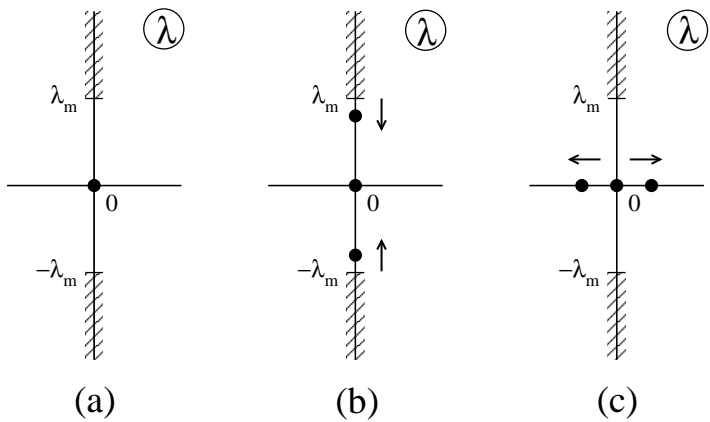}
\caption[]{ \label{fig:imode_collision}
Schematic presentation of the origin of the bifurcation-induced soliton instabilities: 
(a)~spectrum of the integrable cubic NLS model, 
(b)~bifurcation of the soliton internal mode, 
(c)~collision of the internal mode with the neutral mode resulting in the Vakhitov-Kolokolov-type soliton instability}
\end{figure}

\begin{figure}
\includegraphics[width=.8\textwidth]{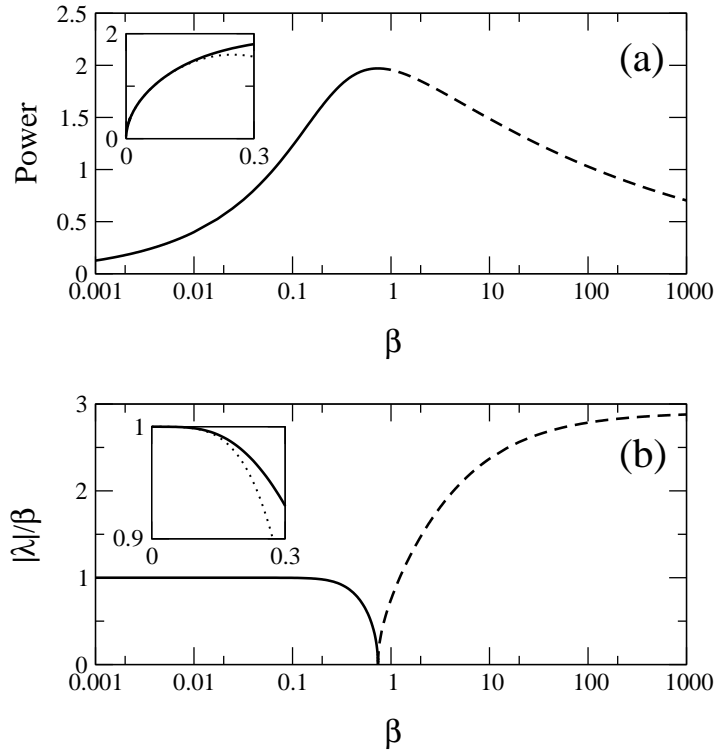}
\caption[]{ \label{fig:power_imode}
Soliton instability in the model (\ref{eq:NLS}),(\ref{eq:perturb_NLS}) and~(\ref{eq:power_perturb}), presented through 
(a)~the power dependence $P(\beta)$ and 
(b)~the evolution of the discrete eigenvalue of the problem (3) that defines the soliton internal mode (solid) and an instability mode (dashed). Solitons for $\beta > \beta_{\rm cr}$, i.e. for $dP/d\beta < 0$ [dashed curves in~(a) and~(b)] are {\em linearly unstable}. Dotted lines show the asymptotic dependences calculated analytically}
\end{figure}

To demonstrate that internal modes are {\em generic for nonintegrable models}, we consider a weakly-perturbed cubic NLS equation with the nonlinear term,
\begin{equation} \label{eq:perturb_NLS}
   {\cal F}(I; x) = I + \epsilon f(I) ,
\end{equation}
where $f(I)$ describes a deviation from the Kerr nonlinear response, and $\epsilon$ is a small parameter. Then, the stationary solution can be expressed asymptotically as $\Phi(x; \beta) = \Phi_0 (x) + \epsilon \Phi_1(x) + {\rm O}(\epsilon^2)$, where $\Phi_0(x) = \sqrt{2 \beta} {\rm sech}\, ( \sqrt{\beta} x )$ is the soliton of the cubic NLS equation, and $\Phi_1(x)$ is a localized correction derived from Eqs.~(\ref{eq:NLS}) and~(\ref{eq:perturb_NLS}). Neglecting the second-order corrections, we find the results for the effective potentials of the linearized eigenvalue problem~(\ref{eq:linear_eigen}), $U_0 = \Phi_0^2 + \epsilon \widetilde{U}_0$ and $U_1 = 3 \Phi_0^2 + \epsilon \widetilde{U}_1$, where
$\widetilde{U}_0 = f(\Phi_0^2) + 4 \Phi_0 \Phi_1$ and
$\widetilde{U}_1 = f(\Phi_0^2) + 2 \Phi_0^2 f^{\prime}(\Phi_0^2) + 12 \Phi_0 \Phi_1$, and the prime denotes differentiation with respect to the argument.

The linear eigenvalue problem~(\ref{eq:linear_eigen}) and~(\ref{eq:perturb_NLS}) can be solved exactly at $\epsilon =0$ (see, e.g., Ref.~\cite{kaup}). Its discrete spectrum contains only the degenerated eigenvalue at the origin, $\lambda = 0$, corresponding to the so-called {\em soliton neutral mode} [see Fig.~\ref{fig:imode_collision}(a)]. We can show that a small perturbation can lead to {\em the creation of an internal mode} (that corresponds to two symmetric discrete eigenvalues), which bifurcates from the continuous spectrum band, as shown in Fig.~\ref{fig:imode_collision}(b). To be definite, we consider the upper branch of the spectrum and suppose that the cut-off frequencies $\lambda_{\rm m} = \pm \beta$ are not shifted by the perturbation. Then, the internal mode frequency can be presented in the form $\lambda = \beta - \epsilon^2 \kappa^2$, where $\kappa$ is defined by the following result~\cite{prl_internal}:
\begin{equation} \label{eq:kap_NLS}
  |\kappa| 
  = \frac{1}{4} {\rm sign} (\epsilon) 
    \int_{-\infty}^{\infty} 
        \left\{   V(x,\beta) \widetilde{U}_1 V(x; \beta) 
                + W(x,\beta) \widetilde{U}_0 W(x; \beta) \right\} dx . 
\end{equation}
Here $\{V(x; \beta), W(x; \beta) \}$
are the eigenfunctions of the cubic NLS equation calculated at the edge of the continuous spectrum, $V(x; \beta) = 1 - 2\; {\rm sech}^2 (\sqrt{\beta} x)$ and $W(x; \beta) = 1$. A soliton  internal mode appears if the right-hand side of Eq.~(\ref{eq:kap_NLS}) is positive.

As an important example, we consider the case of the NLS equation~(\ref{eq:NLS}),(\ref{eq:perturb_NLS}) perturbed by a higher-order power-law nonlinear term, 
\begin{equation} \label{eq:power_perturb}
   f(I) = \epsilon I^{3}.
\end{equation}
The first-order correction to the soliton profile can be found in the form 
\[
  \Phi_1(x) 
  = - \frac{\sqrt{2} \beta^{5/2} 
                     \left[   2 {\rm cosh}(2 \sqrt{\beta} x) 
                            +   {\rm cosh}(4 \sqrt{\beta} x) \right]}{
                     3 {\rm cosh}^{5}(\sqrt{\beta} x)} .
\]
With the help of Eq.~(\ref{eq:kap_NLS}), it is easy to show that for $\epsilon > 0$ a perturbed NLS soliton possesses an internal mode that can be found analytically near the continuum spectrum edge,
\begin{equation}
  \lambda = \beta \left[ 1 - 
              {\left(\frac{ 64 \epsilon }{15}\right)}^2 \beta^4 
              + O( \beta^6 ) \right].
\end{equation}
\begin{figure}
\includegraphics[width=.5\textwidth]{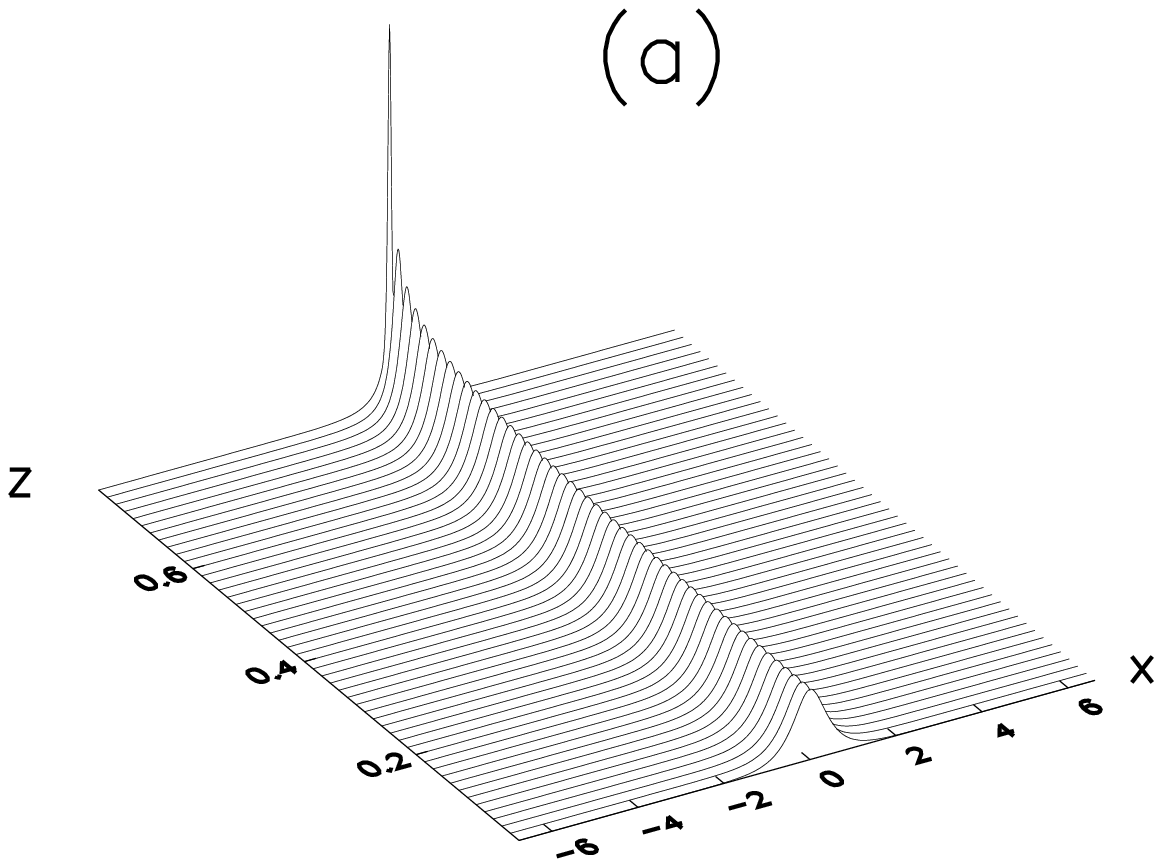}
\includegraphics[width=.5\textwidth]{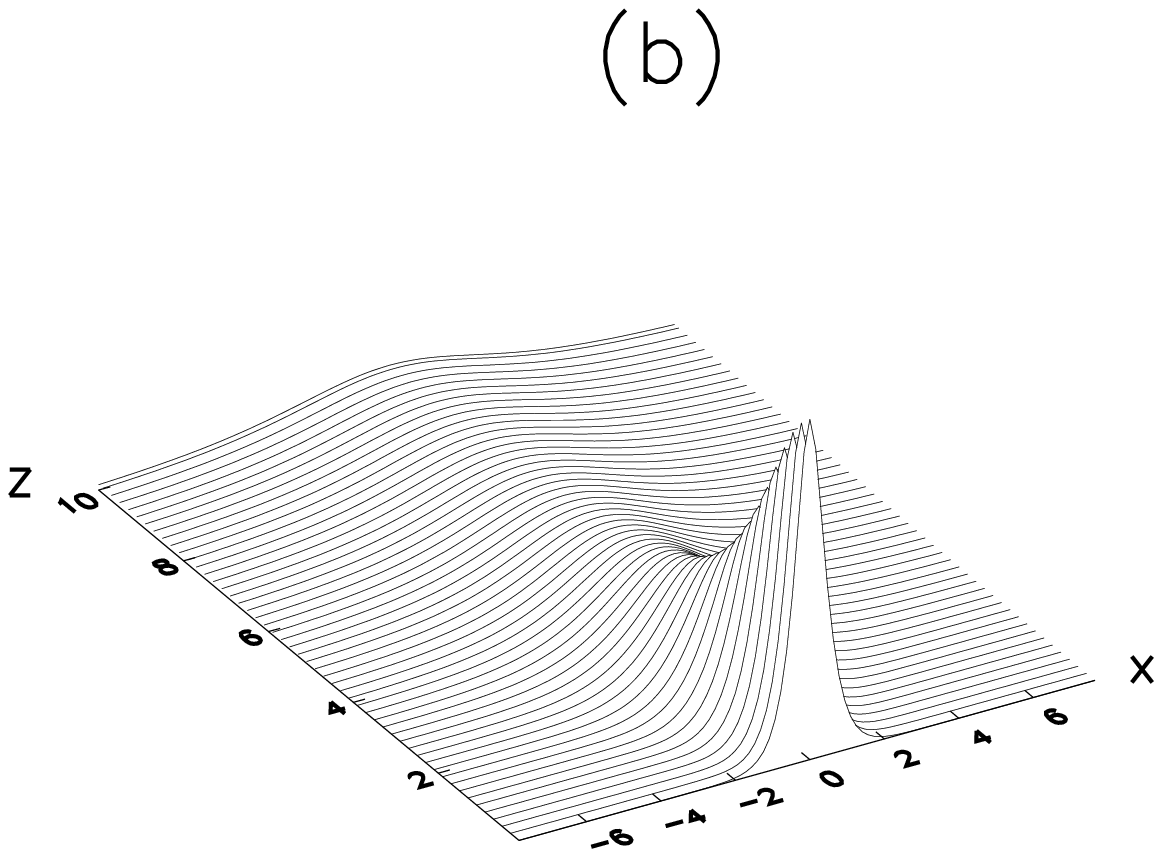}
\caption[]{ \label{fig:bpm_pnls}
Evolution of a perturbed unstable NLS soliton in the model~(\ref{eq:NLS}),(\ref{eq:perturb_NLS}),(\ref{eq:power_perturb}) for  $\epsilon=1$, and $\beta =2$, in the case of  
(a)~increased power (collapse) and 
(b)~decreased power (switching to a low-amplitude stable state). 
The initial power was changed by 1\% compared to the exact soliton solution}
\end{figure}

For high intensities, the additional nonlinear term~(\ref{eq:power_perturb}) is no longer small, and the soliton solutions, together with the associated linear spectrum, should be calculated numerically. Power dependence $P(\beta)$ calculated with the help of Eq.~(\ref{power}) for the soliton of the model~(\ref{eq:NLS}),(\ref{eq:perturb_NLS}), and~(\ref{eq:power_perturb}) is presented in Fig.~\ref{fig:power_imode}(a), and it is matched with the discrete eigenvalue of the linearized problem~(\ref{eq:linear_eigen}) shown in Fig.~\ref{fig:power_imode}(b). First of all, we notice that the asymptotic theory provides accurate results for small intensity solitons, i.e. for $\beta < 0.1$ (shown by the dotted curves in the insets).
Secondly, the slope of the power dependence changes its sign at the point $\beta = \beta_{\rm cr}$, where the soliton internal mode vanishes colliding with the soliton neutral mode, as depicted in Figs.~\ref{fig:imode_collision}\mbox{(b,c)}. At that point, the soliton stability changes due to the appearance of a pair of unstable ({\em purely imaginary}) eigenvalues [dashed curve in Fig.~\ref{fig:power_imode}(b)].  In Sec.~\ref{sec:VK} below, we prove rigorously a link between the soliton stability and the slope of the dependence $P(\beta)$.

For $\beta \approx \beta_{\rm cr}$, the instability-induced dynamics of an unstable soliton can be described by approximate equations for the soliton parameters derived by the multi-scale asymptotic technique (see Sec.~\ref{sec:marginal_stability} below) but, in general,  we should perform numerical simulations in order to study the evolution of linearly unstable solitons. In Figs.~\ref{fig:bpm_pnls}(a,b), we show two different types of the instability-induced soliton evolution in our model.  In the first case, a small perturbation that effectively {\em increases} the soliton power results in an unbounded growth of the soliton amplitude and subsequent beam collapse [see Fig.~\ref{fig:bpm_pnls}(a)]. In the second case, a small {\em decrease} of the soliton power leads to a switching of a soliton of an unstable [dashed, Fig.~\ref{fig:power_imode}(a)] branch to a stable [solid, Fig.~\ref{fig:power_imode}(a)] one, as is shown in Fig.~\ref{fig:bpm_pnls}(b). This latter scenario becomes possible because, in the model under consideration, all small-amplitude solitons are stable.  However, if the  small-amplitude solitons are unstable, the soliton beam does not converge to a stable state but, instead, diffracts. Thus, in the NLS-type nonlinear models there exist {\em three distinct types} of the instability-induced solitons dynamics~\cite{P1}.

\section{Stability criterion for fundamental solitons} \label{sec:VK}

Direct investigation of the eigenvalue problem~(\ref{eq:linear_eigen}) is a complicated task which, in general, does not yield a complete analytical solution. However, for a class of ``fundamental'' solitary waves
(i.e. solitons with no nodes), the analysis can be greatly simplified. First, we reduce the system~(\ref{eq:linear_eigen}) to a single equation:
\begin{equation} \label{eq:L_v1}
  L_0 L_1 v = \lambda^2 v , \nonumber
\end{equation}
for which the stability condition requires all eigenvalues $\lambda^2$ to be positive. It is straightforward to show that $L_0=L^{+} L^{-}$, where $L^\pm = \pm d / dx\; + \Phi^{-1} (d \Phi / d x)$, and thus, instead of~(\ref{eq:L_v1}), one can consider an auxiliary eigenvalue problem~\cite{ltp1},
\begin{equation} \label{eq:L_vt1}
  L^{-} L_1 L^{+} \tilde{v} = \lambda^2 \tilde{v} , \nonumber
\end{equation}
which reduces to Eq.~(\ref{eq:L_v1}) after the substitution $v = L^{+} \tilde{v}$. Since the operator $L^{-} L_1 L^{+}$ is Hermitian,  all eigenvalues $\lambda^2$ of Eqs.~(\ref{eq:L_v1}) and~(\ref{eq:L_vt1}) are real, and in this case {\em oscillatory instabilities do not occur}.

Properties of the operators $L_j$ ($j=0,1$) are well-studied in the literature, in particular,  as a characteristic example of the spectral theory of the second-order differential operators (see, e.g.,  Ref.~\cite{titchmarsh}). For our problem, we use two {\em general mathematical results} about the spectrum of the linear eigenvalue problem $L_j \varphi_n^{(j)} = \lambda_n^{(j)} \varphi_n^{(j)}$: 
\begin{itemize}
\item   the eigenvalues can be ordered as $\lambda_{n+1}^{(j)} > \lambda_n^{(j)}$, where $n \ge 0$ defines the number of zeros in the corresponding eigenfunction  $\varphi_n^{(j)}$; 
\item  for a ``deeper'' potential well, $\widetilde{U}_j(x) \ge U_j(x)$, the 
corresponding set of the eigenvalues is shifted ``down'', 
i.e. $\widetilde{\lambda}_n^{(j)} \le \lambda_n^{(j)}$.
\end{itemize}

Let us first discuss the properties of the operator $L_0$, for which the soliton neutral mode is an eigenstate, i.e. $L_0 \Phi(x; \beta) = 0$. As we have assumed earlier, $\Phi(x; \beta)>0$ is the ground state solution with no nodes and, therefore, $\lambda_n^{(0)} > \lambda_0^{(0)} = 0$ for $n>0$. This means that the operator $L_0$ is positive definite on the subspace of the functions orthogonal to $\Phi(x; \beta)$, which allows to use several general theorems~\cite{VK,ZK,jones,Gr,Gr1} in order to link the soliton stability properties to the number of negative eigenvalues of the operator $L_1$. Specifically, 
\begin{itemize}
\item  the soliton instability appears if there are two (or more) negative eigenvalues, i.e. $\lambda_1^{(1)} < 0$; 
\item the solitons are always stable provided the operator $L_1$ is positively definite;
\item  in an intermediate case, the soliton stability depends on the slope of the power dependence $P(\beta)$, according to the Vakhitov-Kolokolov stability criterion~\cite{VK}, i.e. the soliton is {\em stable} if $\partial P / \partial \beta > 0$, and it is {\em unstable}, otherwise. 
\end{itemize}
Thus, to distinguish between these cases, it is sufficient to determine the signs of the zeroth and the first eigenvalues of the operator $L_1$. 

To demonstrate the validity of the Vakhitov-Kolokolov criterion, we follow the standard procedure~\cite{VK}. First, we note that, for the fundamental solitons with no nodes, both the direct, $L_0$, and inverse, $L_0^{-1}$, operators exist, and they are positively definite for any function orthogonal to $\Phi(x; \beta)$, which can only be a neutral eigenmode of Eq.~(\ref{eq:L_v1}), and therefore can be ignored. Thus, by applying the inverse operator $L_0^{-1}$ to Eq.~(\ref{eq:L_v1}), we obtain another linear problem with the same spectrum:
\begin{equation} \label{eq:L_v2}
  L_1 v = \lambda^2 L_0^{-1} v ,
\end{equation}
where $v(x)$ now satisfies the orthogonality condition:
\begin{equation} \label{eq:uv_ort}
  \left< v | \Phi \right> \equiv 
           \int_{- \infty}^{+ \infty} v^{\ast}(x) \Phi(x; \beta) dx = 0.
\end{equation}

Second, we multiply both sides of Eq.~(\ref{eq:L_v2}) by the function $v^{\ast}(x)$,  integrate over $x$, and obtain the following result,
\begin{equation} \label{eq:omega_min}
  \lambda^2 = \frac{ \left< v | L_1 v \right> }{ 
                                     \left< v | L_0^{-1} v \right>} .
\end{equation}
Because the denominator in Eq.~(\ref{eq:omega_min}) is positively definite for $v$ satisfying Eq.~(\ref{eq:uv_ort}), the sign of this ratio depends only on the numerator.
The instability will appear if there exists an eigenvalue $\lambda^2 < 0$ (so that $\lambda$ is imaginary), and this is possible only if
\begin{equation} \label{eq:L1_min}
  \min \left< v | L_1 v \right>\; < 0 ,
\end{equation}
where we normalize the function $v(x)$ to make the expression in Eq.~(\ref{eq:L1_min}) finite,
\begin{equation} \label{eq:vv1}
 \left< v | v \right> = 1.
\end{equation}

In order to find the minimum in Eq.~(\ref{eq:L1_min}) under the
constraints given by Eqs.~(\ref{eq:uv_ort}) and~(\ref{eq:vv1}), we use the
method of Lagrange multipliers, and look for a minimum of the
following functional:
\[
  {\cal L} = \left< v | L_1 v \right> - \nu \left< v | v \right>
              - \mu \left< v | \Phi \right> ,
\]
where real parameters $\nu$ and $\mu$ are unknown. With no lack of generality, we assume that $\mu \ge 0$, as otherwise the sign of the function $v(x)$ can be inverted. The extrema point of the functional ${\cal L}$ can be then found from the condition $\delta {\cal L} / \delta v = 0$, where $\delta$ denotes the variational derivative. As a result, we obtain the following equation:
\begin{equation} \label{eq:L1_vmin}
  L_1 v  = \nu v + \mu \Phi ,
\end{equation}
where the values of $\nu$ and $\mu$ should be chosen in such a way that the conditions~(\ref{eq:uv_ort}) and~(\ref{eq:vv1}) are satisfied.
Then, it follows from above that $\left< v | L_1 v \right> = \nu \left< v | v \right>$ and, according to Eq.~(\ref{eq:L1_min}), the stationary state is unstable if and only if there exists a solution with $\nu < 0$.

Operator  $L_1$ has a full set of orthogonal eigenfunctions $\varphi_n^{(1)}$~\cite{titchmarsh}, i.e. $\left< \varphi_n^{(1)} | \varphi_m^{(1)} \right> = 0$ if $n \ne m$. The spectrum of $L_1$ consists of discrete ($\lambda_n^{(1)} < \beta$) and continuous ($\lambda_n^{(1)} \ge \beta$) parts, and the norms of the eigenmodes, $\left< \varphi_n^{(1)} | \varphi_n^{(1)} \right>$, are scaled to unity and a delta-function, respectively. Then, we can decompose the function $v(x)$ in the following way:
\begin{equation} \label{eq:v_vn}
  v(x) = \sum_n D_n \varphi_n^{(1)}(x) 
         + \int_{\beta}^{+\infty} D_n \varphi_n^{(1)}(x) d \lambda_n^{(1)},
\end{equation}
where the sum goes only over the eigenvalues of the discrete spectrum of $L_1$. Coefficients in Eq.~(\ref{eq:v_vn}) can be found as
  $D_n = \left< \varphi_n^{(1)} | v \right> $.
Function $\Phi(x; \beta)$ can be decomposed in a similar way, with the coefficients $C_n = \left< \varphi_n^{(1)} | \Phi \right>$. Then, Eq.~(\ref{eq:L1_vmin}) can be reduced to:
\begin{equation} \label{eq:v_Cn}
  D_n = \left\{ \begin{array}{l} 
         \mu \;C_n / ( \lambda_n^{(1)} - \nu ), 
                             \; {\rm if}\; C_n \ne 0 \; 
                                {\rm and}\; \mu > 0, \\
         1,
                             \; {\rm if}\; C_n = 0,\; \mu = 0,\;
                                {\rm and}\; \nu = \lambda_n^{(1)}, \\
         0,
                             \; {\rm otherwise } .
   \end{array} \right.
\end{equation}
Note that we should not consider degenerate solutions $v \equiv 0$, and therefore $\mu=0$ is only possible if $\nu = \lambda_n^{(1)}$ and $C_n=0$. In order to find the Lagrange multiplier $\nu$, we substitute 
Eqs.~(\ref{eq:v_vn}) and~(\ref{eq:v_Cn}) into the orthogonality condition~(\ref{eq:uv_ort}), and obtain the following equation for the
parameter $\nu$:
\begin{equation} \label{eq:lambda}
 Q(\nu) \equiv \left< v | \Phi \right>
 = \sum_n C_n D_n^{\ast} 
    + \int_{\beta}^{+\infty} C_n D_n^{\ast} d \lambda_n^{(1)} 
 = 0 .
\end{equation}
As has been mentioned above, instability appears if there exists a root 
$\nu < 0$, and thus we should determine the sign of minimal $\nu$ solving Eq.~(\ref{eq:lambda}). Because the lowest-order modes of the operators $L_0$ and $L_1$, $\Phi(x; \beta)$ and $\varphi_0^{(1)}$ respectively, do not contain zeros, the coefficient $C_0 \ne 0$. Then, from the structure of Eq.~(\ref{eq:lambda}) it follows that $Q (\nu < \lambda_0^{(1)} ) > 0$, and thus the solutions are only possible for $\nu > \lambda_0^{(1)}$, meaning that if $\lambda_0^{(1)} \ge 0$ the stationary state $\Phi(x; \beta)$ is {\em stable}.

We notice that the function $Q(\nu)$ is monotonic in the interval $(-\infty,+\infty)$ for $\mu>0$ and $\lambda_0^{(1)} < \nu < \lambda_n^{(1)}$, where $n \ge 1$ corresponds to the smallest eigenvalue with $C_n \ne 0$. Then, it immediately follows that instability appears if $\lambda_1^{(1)}<0$ and $C_1 \ne 0$. On the other hand, if $C_1 = 0$, the corresponding eigenmode $\varphi_1^{(1)}$ satisfies Eq.~(\ref{eq:L1_vmin}) and the constraints~(\ref{eq:uv_ort}),(\ref{eq:vv1}) with $\nu = \lambda_1^{(1)}$ and $\mu=0$ and, therefore, an instability is always present if $\lambda_1^{(1)} < 0$.

The last possible scenario is when \mbox{$\lambda_0^{(1)} < 0$} and
\mbox{$\lambda_1^{(1)} \ge 0$}, so that the modes with \mbox{$\nu = \lambda_n^{(1)}$} do not give rise to instability, and we search for the solutions with  $\mu > 0$. Then, because $\lambda_n^{(1)} > \lambda_1^{(1)} \ge 0$, the sign of the solution $\nu$ is determined by the value of $Q(0)$. Indeed, if $Q(0) > 0$, the function $Q(\nu)$ vanishes  at some $\nu < 0$ which indicates {\em instability}, and vise versa. From Eqs.~(\ref{eq:L1_vmin}) and~(\ref{eq:lambda}),  it follows that $Q(\nu=0) = \left< L_1^{-1} \mu \Phi | \Phi \right>$. To calculate this value, we differentiate the equality $L_0 \Phi = 0$ with respect to the propagation constant and obtain
\begin{equation} \label{eq:L1_uk}
 L_1 \frac{\partial \Phi}{\partial \beta} = - \Phi ,
\end{equation}
that finally gives: ${\rm sign}[Q(0)] = - {\rm sign} (d P / d \beta)$. These results provide a proof of the stability conditions outlined above.

An important case is the soliton stability in a homogeneous medium, when ${\cal F}(I)$ does not depend on $x$. Then, a fundamental soliton has a symmetric profile with a single maximum, and $d \Phi / d x$ is the first-order neutral mode of the operator $L_1$, i.e. $\lambda_1^{(1)} = 0$. Therefore, in such a case the  soliton stability follows directly from the slope of the dependence $P(\beta)$.

Linear stability discussed above should be compared with a more {\em general Lyapunov stability theorem} which states that in a conservative system a stable solution (in the Lyapunov sense) corresponds to an extrema point of an invariant such as the system Hamiltonian, provided it is bounded from below (or above). For the NLS equation, this means that a soliton solution is a stationary point of the Hamiltonian $H$ for a fixed power $P$, and it is found from the variational problem $\delta (H + \beta P) =0$. In order to prove the Lyapunov stability, one needs to demonstrate that, for a class of spatially localized solutions, the soliton realizes a minimum of the Hamiltonian when $P$ is fixed. This fact can be shown rigorously for a homogeneous medium with cubic nonlinearity~\cite{ZK}, i.e. for ${\cal F}(I; x) = I$, and it follows from the integral inequality,
\begin{equation} \label{eq:cond}
  H > H_s + (P^{1/2} - P_s^{1/2}),
\end{equation}
where the subscript ``s'' defines the integral values calculated for the  NLS soliton. Condition~(\ref{eq:cond}) proves the soliton stability for both small and finite-amplitude perturbations, and a similar relation can be derived for generalized nonlinearity, being consistent with the Vakhitov-Kolokolov criterion (see details in Ref.~\cite{ZK}).

\section{Marginal stability point: Asymptotic analysis} 
         \label{sec:marginal_stability}

As follows from the linear stability analysis, the solitons in a homogeneous medium are unstable when the slope of the power dependence is negative, i.e. for $d P / d \beta < 0$. Near the marginal stability point $\beta = \beta_{\rm cr}$ defined by the condition $(dP/d\beta)_{\beta=\beta_{\rm cr}} =0$, where the instability growth rate is small, we can derive {\em a general analytical asymptotic model} which describes not only linear instabilities but also the nonlinear long-term evolution of unstable solitons. Such an approach is based on a nontrivial modification of the soliton perturbation theory~\cite{kivmal}.
Indeed, the standard soliton perturbation theory is usually applied to analyse the soliton dynamics under the action of external perturbations. Here we should deal with {\em a qualitatively different physical problem}  when an unstable bright soliton evolves under the action of its ``own'' perturbations. As a result of the development of such an instability, the soliton propagation constant varies slowly along the propagation direction, i.e. $\beta = \beta(z)$. As the instability growth rate is small near the threshold $\beta = \beta_{\rm cr}$, we can assume that the profile of the perturbed soliton is slowly varying with $z$ and the soliton evolves almost adiabatically (i.e., it remains self-similar). Therefore, we can develop {\em an asymptotic theory} representing the solution to the original model~(\ref{eq:NLS}) in the form $\psi = \phi(x;\beta;Z) \exp [i \beta_0 z + i \epsilon \int_{0}^{Z} \beta(Z^{\prime}) dZ^{\prime}]$, where $\beta = \beta_0 + \epsilon^2 \Omega(Z)$, $Z = \epsilon z$, and $\epsilon \ll 1$. Here the constant value $\beta_0$ is chosen in the vicinity of the critical point $\beta_{\rm cr}$. Then, using the asymptotic multi-scale expansion in the form $\phi = \Phi(x;\beta) + \epsilon^3 \phi_{3}(x;\beta;Z) + O(\epsilon^4)$, we obtain the following equation for the soliton propagation constant $\beta$ (details can be found in Refs.~\cite{P1,MOI}),
\begin{equation} \label{3.1} 
   M (\beta_{\rm cr}) \frac{d^2 \Omega}{d Z^2} 
   + \frac{1}{\epsilon^2} 
     \frac{d P}{d \beta} \biggr|_{\beta =  \beta_0} \Omega + \frac{1}{2} 
     \frac{d^2 P}{d \beta^2} \biggr|_{\beta = \beta_{\rm cr}} \Omega^2   =0, 
\end{equation} 
where $P(\beta)$ and $M(\beta)$ are calculated through the stationary soliton solution, and
\[
M(\beta) = \int_{-\infty}^{+\infty} \left[ \frac{1}{\Phi (x; \beta)} \int_{0}^{x} \Phi (x^{\prime}; \beta) \; \frac{\partial \Phi (x^{\prime}; \beta)}{\partial \beta} dx^{\prime} \right]^2 dx > 0.
\]

A remarkable result which follows from this asymptotic analysis is the following. In the generalized NLS equation~(\ref{eq:NLS}) the dynamics of solitary waves near the marginal stability point $\beta = \beta_{\rm cr}$ can be described by a simple collective-coordinate model~(\ref{3.1}) which is equivalent to the equation of motion of an effective ({\em inertial and conservative}) particle of the mass $M (\beta_{\rm cr})$ with the coordinate $\Omega$ moving under the action of a potential force proportional to the difference $P_0 - P(\beta)$, where $P_0 = P(\beta_0)$.

First two terms in Eq.~(\ref{3.1}) give the result of the linear stability theory, according to which the soliton is {\em linearly unstable} provided $d P / d \beta < 0$. Nonlinear term in Eq.~(\ref{3.1}) allows to describe not only linear but also long-term nonlinear dynamics of an unstable soliton and, moreover, to identify qualitatively different  scenarios of the instability-induced soliton dynamics near the marginal stability point, as discussed in Refs.~\cite{P1,MOI}.

\section{Nonlinear guided waves and impurity modes} \label{sec:nlgw}
\subsection{Model and general remarks}    \label{sec:nlgs_model}

In this section, we demonstrate how the basic concepts and results of the soliton stability theory can be employed to study the stability of nonlinear guided waves in inhomogeneous media.
In particular, we consider the model~(\ref{eq:NLS}) with a real function ${\cal F}(I; x)$ that describes both {\em nonlinear} and {\em  inhomogeneous} properties of an optical medium. This kind of problems has a number of important physical applications ranging from the nonlinear dynamics of solids to the theory of nonlinear photonic crystals and waveguide arrays in optics. In application to the theory of guided electromagnetic waves, the model we discuss below describes a special case of a stratified (or layered) dielectric medium for which {\em nonlinear guided waves} and their stability have been analyzed during the last 20 years~\cite{optics,ms}.

Following the original study presented in Ref.~\cite{impurity_modes}, we consider a simplified case when the inhomogeneity is localized in a small region, and it is created by a thin layer embedded into a nonlinear medium.  Then, if the corresponding wavelength is much larger than the layer thickness, in the continuum-limit approximation the layer response can be described  by a delta-function and, therefore, we can write 
\begin{equation} \label{eq:def_F}
      {\cal F}(I; x) = F(I) + \delta(x) G(I), 
\end{equation}
where the functions $F(I)$ and $G(I)$ characterise the properties of a bulk medium and the layer, respectively. The model~(\ref{eq:NLS}),(\ref{eq:def_F}) describes, in particular, a special case of a more general problem of the existence and stability of nonlinear guided waves in a stratified medium with a Kerr or non-Kerr nonlinear response (see, e.g., some examples in Refs.~\cite{optics}).

When a thin layer is introduced into the system, the translational invariance of the model is broken at the point of the layer location, i.e. at $x=0$. Nonlinear modes of such an inhomogeneous model should be found as spatially localized solutions of the following equation,
\begin{equation} \label{eq:u0_inh}
  - \beta \Phi
  + \frac{d^2 \Phi}{d x^2} 
  + {\cal F}(\Phi^2; x) \Phi 
  = 0 .
\end{equation}
We notice that for the problem at hand the profiles of localized modes should satisfy {\em a homogeneous equation} on both sides of the layer and, therefore, they can be constructed by using the solutions of Eq.~(\ref{eq:u0_inh}) with ${\cal F}(I; x) = F(I)$.
Due to a translational symmetry, such a localized solution
can be presented in the form $\Phi_0(x - x_0)$, where $x_0$ is an arbitrary shift. We also note that the profile is symmetric, $\Phi_0(x) = \Phi_0(-x)$, it does not contain zeros, $\Phi_0(x)>0$, and it has a single hump since $d \Phi_0 / d |x| < 0$ at $x \ne 0$. 
Therefore, a general solution of inhomogeneous Eq.~(\ref{eq:u0_inh}) for a spatially localized normal mode can be presented in the following form,
\begin{equation} \label{eq:u_slv}
   \Phi(x) = \left\{ \begin{array}{l}
                    \Phi_0( x - x_0 ) , \; x \ge 0 , \\
                    \Phi_0( x - s x_0 ) , \; x \le 0 , 
          \end{array} \right.
\end{equation}
where $x_0$ and $s$ are defined through the layer parameters. From the continuity condition at $x=0$, it follows that $s = \pm 1$. Thus, the parameter $s$ defines the type of symmetry of the localized mode, i.e. the mode is {\em symmetric}, for $s = -1$, and {\em asymmetric}, for $s = +1$. Parameter $x_0$ is defined by the following matching condition, 
\begin{equation} \label{eq:x0}
   (1-s) \frac{d \Phi_0}{d x} (x_0) = G(I_0) \Phi_0(x_0),
\end{equation}
where $I_0 = \Phi_0^2(x_0)$.
We note that the power of the asymmetric modes is the same as that for solitons in a homogeneous medium, $P_0 = \int_{-\infty}^{+\infty} \Phi_0^2(x) dx$, while the power of symmetric modes (hereafter denoted as $P$) is smaller than $P_0$ if $x_0 <0$, and larger otherwise.

As follows from Eq.~(\ref{eq:u_slv}), the function $\Phi(x)$ is sign definite. Then, in order to apply the general stability results derived for the fundamental solitons of the NLS-type equations in Sec.~\ref{sec:VK}, we should study the spectral properties of the linear operator $L_1$. Due to the presence of a layer, the number of negative eigenvalues of the operator $L_1$ depends on the mode parameter $\beta$. As has been demonstrated in Ref.~\cite{ms}, if an eigenvalue passes though zero and changes its sign, the so-called {\em critical point} appears in 
the power dependence $P(\beta)$ calculated for a localized mode. One type of such critical points is a {\em bifurcation}, where the mode belongs simultaneously to the two families of localized solutions, symmetric and asymmetric ones. Thus, the stability properties of these distinct types of nonlinear localized modes can be expected to be different, and we discuss these two cases separately.

\subsection{Stability of symmetric and asymmetric modes}
            \label{sec:nlgw_stability}

{\em Symmetric modes}. In this case it can be verified that, similar to the solitary waves of homogeneous models (see Sec.~\ref{sec:VK} above), the operator $L_1$ has a neutral mode $d \Phi / d |x|$ with zero eigenvalue, but only for a special value of the layer response:
\begin{equation} \label{eq:M1_cr}
  G_1^{\rm cr} = \left. 
                   2 \left( \frac{d \Phi_0(x)}{dx} \right)^{-1} 
                     \frac{d^2 \Phi_0(x)}{d x^2} 
                 \right|_{x=x_0} ,
\end{equation}
where $G_1 = G(I_0) + 2 I_0 [d G / d I]_{I=I_0}$. For $x_0>0$ the function $\Phi(x)$ has two humps, and its derivative $d \Phi / d |x|$ corresponds to the second-order mode with $\lambda_2^{(1)} = 0$. On the other hand, for $x_0<0$, the neutral mode describes a ground-state eigenfunction with $\lambda_0^{(1)} = 0$. From the spectral theorem~\cite{titchmarsh}, it follows that this eigenvalue increases for $G_1 < G_1^{\rm cr}$, and decreases otherwise. Additionally, we note that due to the symmetry of the potentials in the linear eigenvalue problem, $U_j(x) = U_j(-x)$, the amplitude of the first-order eigenmode vanishes at the layer location ($x=0$), and thus its eigenvalue $\lambda_1^{(1)}$ does not depend on $G_1$. However, the inequality $\lambda_0^{(0)} < \lambda_0^{(1)} < \lambda_0^{(2)}$ should be fulfilled for any $G_1$, and we immediately come to the conclusion that $\lambda_1^{(1)} < 0$, if $x_0 > 0$, and $\lambda_1^{(1)} > 0$, if $x_0 < 0$.  It is also straightforward to verify that for $x_0 = 0$ we should have $G(I_0) = 0$, and the derivative $d \Phi(x) / d x$, which has a single zero at $x=0$, is a neutral mode of the linear operator $L_1$, i.e. $\lambda_1^{(1)} = 0$. In Table~\ref{tab:stab_symm} we summarize the properties of the operator $L_1$ and the corresponding general conditions for the stability of {\em symmetric nonlinear localized modes}.

\begin{table}
\caption{ Conditions for stability of symmetric modes }
\renewcommand{\arraystretch}{1.4}
\renewcommand{\tabcolsep}{5mm}
\begin{tabular}{@{}lll@{}} 
     Conditions
   & $L_1$ spectrum
   & Stability 
\\ \hline \hline
     $x_0 > 0$ 
   & $\lambda_0^{(1)} < \lambda_1^{(1)} < 0$
   & unstable
\\ \hline
     $x_0 \le 0$, 
     $G_1 \le G_1^{\rm cr}$
   & $0 \le \lambda_0^{(1)} < \lambda_1^{(1)}$
   & stable
\\ \hline 
     $x_0 \le 0$,
     $G_1 > G_1^{\rm cr}$,
     $\partial P / \partial \beta > 0$
   & $\lambda_0^{(1)} < 0 \le \lambda_1^{(1)}$
   & stable
\\ \hline 
     $x_0 \le 0$,
     $G_1 > G_1^{\rm cr}$,
     $\partial P / \partial \beta \le 0$
   & $\lambda_0^{(1)} < 0 \le \lambda_1^{(1)}$
   & unstable
\end{tabular}
\label{tab:stab_symm}
\end{table}

The next important step is to connect the spectral characteristics of the 
linear operator $L_1$ (and thus the stability of nonlinear localized modes)
with the power dependence $P(\beta)$. First, we notice that at the {\em bifurcation point} defined as $P(\beta) = P_0(\beta)$, the parameter $x_0$ passes through zero and the eigenvalue $\lambda_1^{(1)}$ changes its sign (see Table~\ref{tab:stab_symm}). In particular, for $P(\beta) > P_0(\beta)$ such a mode should have {\em a double-hump profile} and, therefore, it is {\em always unstable}. In contrast to these unstable double-hump modes, a family of asymmetric modes emerges at the bifurcation point.

{\em Asymmetric modes.}
Although  asymmetric modes have the profiles of solitons, the corresponding spectrum of the operator $L_1$ can be  different.  Only in the special case $G_1=0$, there exists the first-order neutral mode with $\lambda_1^{(1)} = 0$, i.e. $L_1 (d \Phi / d x) = 0$.

There can exist several families of asymmetric modes, each of them characterized by the constant intensity at the layer $I_0$, so that ${\rm sign}(G_1) = {\rm sign}[d G / d I]_{I=I_0}$ at the point where the layer response vanishes, $G(I_0) = 0$. If $G_1$ is negative, the stability of the localized mode is the same as the stability of a soliton in a bulk medium.  On the other hand, the modes corresponding to positive $G_1$
are unstable, and they can be attracted to or repelled from the nonlinear layer. Performing the analysis similar to that in Ref.~\cite{ms}, we find that for the asymmetric modes $G_1>0$, if after the bifurcation point the branch of the symmetric modes is lower than that of the asymmetric modes, $P(\beta) < P_0(\beta)$. In the opposite case, we have $G_1 < 0$. In Table 2 we summarize the stability conditions for {\em asymmetric nonlinear localized modes}.

\begin{table}
\caption{ Conditions for stability of asymmetric modes }
\renewcommand{\arraystretch}{1.4}
\renewcommand{\tabcolsep}{5mm}
\begin{tabular}{@{}lll@{}} 
     Conditions
   & $L_1$ spectrum
   & Stability 
\\ \hline \hline
     $G_1 > 0$
   & $\lambda_0^{(1)} < \lambda_1^{(1)} < 0$
   & unstable
\\ \hline
     $G_1 \le 0$,
     $\partial P_0 / \partial \beta > 0$
   & $\lambda_0^{(1)} < 0 \le \lambda_1^{(1)}$
   & stable
\\ \hline 
     $G_1 \le 0$,
     $\partial P_0 / \partial \beta \le 0$
   & $\lambda_0^{(1)} < 0 \le \lambda_1^{(1)}$
   & unstable
\end{tabular}
\label{tab:stab_asymm}
\end{table}

\subsection{Example: Competing power-law nonlinearities} 
            \label{sec:nlgw_powernl}

To discuss a characteristic example of the stability theory briefly outlined above, we consider the power-law nonlinear response of both the bulk medium and the layer,
\begin{equation}
  F(I) = I^{\sigma}, \;\;\; G(I) = a + b I^{\gamma},
\end{equation}
where the nonlinearities are characterized by the positive 
powers $\sigma$ and $\gamma$, and the coefficients $a$ and $b$ account for the linear and nonlinear properties of the layer, respectively. 

\begin{figure}
\includegraphics[width=.8\textwidth]{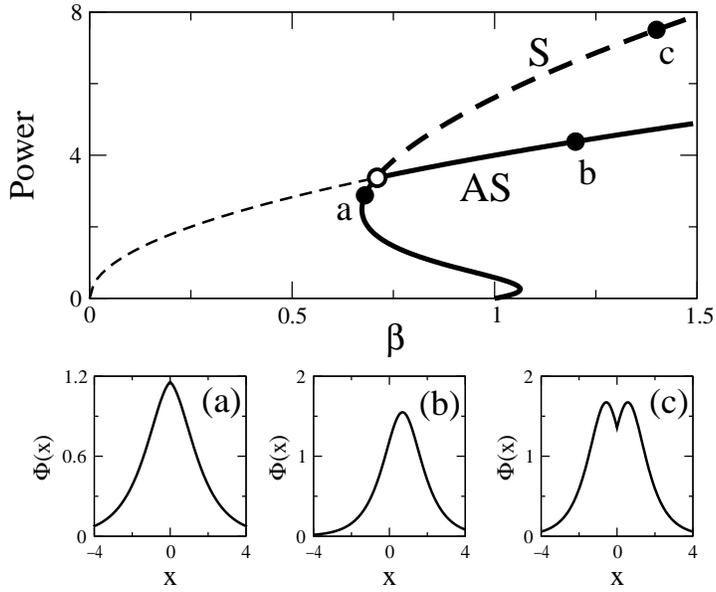}
\caption{ \label{fig:pwr-sf-al_p-beta_m_s1g2} 
Power vs. propagation constant diagram for $\sigma=1$, $\gamma=2$, $a=2$, and $b=-1$.
Solid and dashed parts of the main curves correspond to
stable and unstable localized modes, respectively.
Open circle~--- the bifurcation point from the 
symmetric (S) to asymmetric (AS) modes.
Thin dashed~--- the soliton power $P_0(\beta)$. 
The modes corresponding to the marked points (a),(b), and~(c) are shown below}
\end{figure}

We consider the case when the layer itself possesses {\em repulsive nonlinearity} (i.e., $b < 0$). Due to a competition of localizing and de-localizing nonlinearities, the power dependence of the family of localized {\em symmetric} modes can become {\em multi-valued}, i.e. two or three states may correspond to the same value of the mode propagation constant (see Fig.~\ref{fig:pwr-sf-al_p-beta_m_s1g2}). Moreover, such states can exist below and above the linear cut-off.

\begin{figure}
\sidecaption
\includegraphics[width=.6\textwidth]{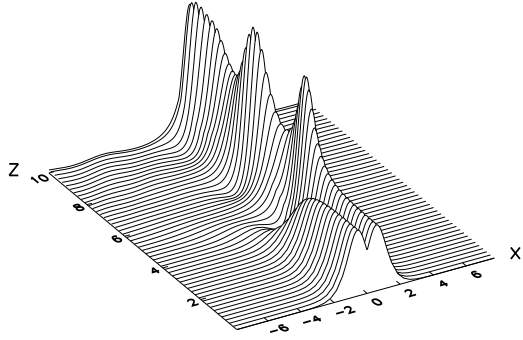}
\caption{ \label{fig:bpm-sf-al_p-beta_m}
Evolution of a slightly perturbed symmetric localized mode shown in Fig.~\ref{fig:pwr-sf-al_p-beta_m_s1g2}(c).
An unstable two-hump mode eventually transforms into a stable asymmetric mode}
\end{figure}

The branch of the asymmetric solutions emerges at the bifurcation point, $\beta_b$, where the power of the symmetric localized modes, $P(\beta_b)$, coincides with the power of solitons in a homogeneous medium, $P_0(\beta_b)$ (open circle in Fig.~\ref{fig:pwr-sf-al_p-beta_m_s1g2}). Stability of asymmetric modes for $b < 0$ is the same as that of the solitons in a bulk medium, i.e. the modes are stable for $\sigma < 2$ and unstable otherwise. On the other hand, after the bifurcation point, i.e. for $\beta > \beta_b$, the symmetric localized mode becomes two-humped and, therefore, it is unstable to asymmetric perturbations.
In Fig.~\ref{fig:bpm-sf-al_p-beta_m}, we illustrate the evolution of a two-hump symmetric mode when the power of a bulk nonlinearity is below the instability threshold ($\sigma < 2$). 
In this case, the symmetry-breaking instability is observed, and the mode is transferred to one side of the layer. Such an excitation of a stable asymmetric state is accompanied by a slowly decaying quasi-periodic oscillations of the mode amplitude. 

\section{Multi-component solitary waves} \label{sec:multi_comp}
\subsection{General theory: $N$ coupled NLS equations}
            \label{sec:multi_comp_general}

As has been already mentioned in the Introduction, the Vakhitov-Kolokolov stability criterion is valid for one-parameter solitary waves, such as scalar non-Kerr solitons, two-wave parametric solitary waves in a quadratic medium, etc. As we demonstrated in Sec.~\ref{sec:nlgw}, the Vakhitov-Kolokolov criterion can also describe, in some cases, the stability of nonlinear guided waves in a scalar model. However, when solitary waves described by nonintegrable nonlinear models possess many parameters, the stability analysis becomes more involved, and in many cases a direct numerical study of the corresponding eigenvalue problem is required. Nevertheless, in some cases it is possible to extend the stability analysis to multi-parameter solitary waves. Below, we follow the recent original work by Pelinovsky and Kivshar~\cite{rigorous} and demonstrate how the multi-parameter generalization of the Vakhitov-Kolokolov stability criterion can be derived for a system of $N$ coupled NLS equations. Such a theory includes, as a limiting case, the bifurcation analysis near the marginal stability curve. 

We consider a system of $N$ incoherently coupled NLS equations,
\begin{equation} \label{NLS}
  i \frac{\partial \psi_{n}}{\partial z} 
  + d_n \nabla^2_{\bf x} \psi_{n}
  + \left( \sum_{m=1}^N \gamma_{nm} |\psi_m|^2 \right) \psi_n 
  = 0,
\end{equation}
where $\nabla_{\bf x}^2$ stands for Laplacian in the $D-$dimensional space ${\bf x} = (x_1,...,x_D)$, and all the coefficients $d_n$ are assumed to be positive. When one of the variables of the vector ${\bf x}$ stands for time, Eqs.~(\ref{NLS}) describe the spatio-temporal dynamics of self-focused and self-modulated light in  the form  of the so-called {\em light bullets}.

Provided the symmetry conditions $\gamma_{nm} = \gamma_{mn}$ are satisfied, the system~(\ref{NLS}) conserves the Hamiltonian,
\[
   H 
   = \int_{-\infty}^{\infty} d {\bf x} 
        \left(   \sum_{n=1}^N d_n \left| \nabla_{\bf x} \psi_{n} \right|^2 
               - \frac{1}{2} \sum_{n=1}^N \sum_{m=1}^N 
                                \gamma_{nm} |\psi_n|^2 |\psi_m|^2 \right),
\]
the individual mode powers,
$P_n = \frac{1}{2} \int |\psi_n|^2 d {\bf x}$,
and the total field momentum.  Localized solutions of Eqs.~(\ref{NLS})
for the multi-component fundamental solitary waves are defined in a standard form as
$\psi_n = \Phi_n({\bf x}) e^{i \beta_n z}$,  where $\Phi_n({\bf x})$
are real functions {\em with no nodes},  and $\beta_n$ are the
{\em positive} propagation constants. Such localized solutions can be also obtained as  stationary points of the Lyapunov functional,
\begin{equation} \label{Lyp}
  \Lambda[\mbox{\boldmath $\psi$}] 
  = H[\mbox{\boldmath $\psi$}] 
  + \sum_{n=1}^N \beta_n P_n[\mbox{\boldmath $\psi$}],
\end{equation}
and, therefore, the first variation of $\Lambda$ vanishes, whereas the second variation 
$\delta^2 \Lambda [\psi]$ defines the stability properties: the negative directions of
the second variation correspond to unstable eigenvalues
in the soliton stability problem.

The stability problem is defined by minimizing the second variation
of the Lyapunov functional $\Lambda[\mbox{\boldmath $\psi$}]$,
\begin{equation} \label{second}
 \delta^2 \Lambda 
 = \int_{-\infty}^{\infty} d {\bf x}
     \left[   \langle {\bf u} | {\bf L}_1 {\bf u} \rangle 
            + \langle {\bf w} | {\bf L}_0 {\bf w} \rangle \right],
\end{equation}
where ${\bf u}({\bf x})$ and ${\bf w}({\bf x})$ are perturbations
of the multi-component solitary wave taken in the form $\mbox{\boldmath $\psi$} = {\bf \Phi}({\bf x}) + \left[ {\bf u} + i {\bf w} \right]({\bf x}) e^{\lambda z}$, and the scalar product is defined as $\langle {\bf f} | {\bf g} \rangle = \sum_{n=1}^N f_n^* g_n$. The matrix Sturm-Liouville operator ${\bf L}_0$ has a diagonal form with the elements
$$
(L_0)_{nn} = - d_n \nabla^2_{\bf x} + \beta_n
- \sum_{m=1}^N \gamma_{nm} \Phi_m^2,
$$
and the matrix operator ${\bf L}_1$ has the elements
$$
(L_1)_{nn} = - d_n \nabla^2_{\bf x} + \beta_n
- \sum_{m=1}^N \gamma_{nm} \Phi_m^2 - 2 \gamma_{nn} \Phi_n^2,
$$
at the diagonal, and $(L_1)_{nm} = - 2 \gamma_{nm} \Phi_n \Phi_m$,
off the diagonal. Similar to the stability problem for one-component solitary waves discussed in Secs.~\ref{sec:linear_eigen} and~\ref{sec:VK}, the matrix operators $L_0$ and $L_1$ define the linear eigenvalue problem for stability of multi-component solitary waves,
\begin{equation} \label{eigenvalue}
  {\bf L}_1 {\bf u} = - \lambda {\bf w}, \;\;\;\;
  {\bf L}_0 {\bf w} =   \lambda {\bf u}.
\end{equation}
Both the linear problem~(\ref{eigenvalue}) and minimization
problem~(\ref{second}) should satisfy a set of $N$ additional constraints,
\begin{equation} \label{constr}
  F_n 
  = \int_{-\infty}^{\infty} d {\bf x}
                \langle \Phi_n {\bf e}_n | {\bf u} \rangle 
  = 0,
\end{equation}
where ${\bf e}_n$ is the $n^{\mbox{th}}$ unit vector, which correspond to
the conservation of the individual powers $P_n$ under the action of a vector perturbation (${\bf u},{\bf w}$).

First of all, we recall that the one-parameter solitary waves with no nodes ($N=1$) are stable in the framework of the constrained variational problem
(\ref{second})-(\ref{constr}) provided the energetic surface
$\Lambda_s(\mbox{\boldmath $\beta$}) = \Lambda[{\bf \Phi}]$ is concave up, i.e.
\begin{equation} \label{crit}
   \frac{d^2 \Lambda_s}{d \beta_1^2} 
   = \frac{d P_1}{d \beta_1} 
   > 0.
\end{equation}
Under this condition, the linear eigenvalue problem~(\ref{eigenvalue})
{\em has no unstable eigenvalues}, i.e. those with  {\em a positive real
part} $\lambda$. Otherwise, the second variation~(\ref{second}),  constrained by the $N$ conditions~(\ref{constr}),  has {\em a single negative  direction} that corresponds to a single positive eigenvalue $\lambda$ in the linear eigenvalue problem~(\ref{eigenvalue}) (see Refs.~\cite{VK,ZK} and discussions above). The stability criterion for scalar (or one-component) NLS solitons holds when the self-adjoint operator ${\bf L}_1$ has a single negative eigenvalue, i.e. when the second variation~(\ref{second}), without the constraint~(\ref{constr}) imposed, has a single negative direction. If the latter condition is not satisfied,  as it happens for solitary waves with nodes, the fundamental criterion for the soliton instability can be extended only for special cases,  while more generic mechanisms of oscillatory instabilities, associated with complex eigenvalues of the linear eigenvalue problem,  may appear beyond the prediction of the fundamental criterion. 

In some particular cases, the soliton stability analysis can be extended to
multi-component solitary waves. In particular, this is possible for a system of incoherently coupled NLS equations~(\ref{NLS}). We assume that the number of negative directions (eigenvalues) of the second variation $\delta^2 \Lambda$ is fixed, and we denote it as $n(\Lambda)$. The unstable eigenvalues $\lambda$ of the linear problem~(\ref{eigenvalue}) are connected with some negative eigenvalues of the matrix ${\bf U}$ defined by the elements
\begin{equation} \label{Hessian}
   U_{nm} 
   = \frac{\partial^2 \Lambda_s}{\partial \beta_n \partial \beta_m}
   = \frac{\partial P_n}{\partial \beta_m} 
   = \frac{\partial P_m}{\partial \beta_n}.
\end{equation}
The matrix ${\bf U}$ is {\em the Hessian matrix} of the soliton energy surface $\Lambda_s(\mbox{\boldmath $\beta$})$. We denote the number of positive eigenvalues of the matrix ${\bf U}$ as $p(U)$, and the number of its negative eigenvalues as $n(U)$, so that $p(U) + n(U) \leq N$, since some eigenvalues may be zeros in a degenerate (bifurcation) case.  As is shown in Ref.~\cite{rigorous}, both $p(U)$ and $n(U)$ satisfy some additional constraints,
\begin{equation} \label{restrict}
   p(U) \leq \min\{N,n(\Lambda)\}, \;\;\;\;
   n(U) \geq \max\{0,N-n(\Lambda)\}.
\end{equation}
Within these notations, the following results on stability and instability of multi-component fundamental solitary waves of the coupled NLS equations~(\ref{NLS}) can be formulated and proved~\cite{rigorous}:

\begin{itemize}

\item  the linear problem~(\ref{eigenvalue}) may have maximum $n(\Lambda)$
unstable eigenvalues $\lambda$, all {\em real} and {\em positive}; 

\item  a multi-component fundamental soliton is {\em linearly unstable} provided $p(U) < n(\Lambda)$; then the linear problem~(\ref{eigenvalue}) has $n(\Lambda) - p(U)$ real (positive or zero-becoming-positive) eigenvalues $\lambda$; 

\item  a multi-component fundamental soliton is {\em linearly stable} provided $p(U) = n(\Lambda) (\leq N)$; in the case  $n(\Lambda) = N$ this criterion implies that the energetic surface  $\Lambda_s(\mbox{\boldmath $\beta$})$ is concave up in the $\mbox{\boldmath $\beta$}$-space; 

\item  a single eigenvalue $\lambda$ crosses {\em a marginal stability curve} when the matrix ${\bf U}$ possesses a zero-becoming-negative eigenvalue; the normal form for the instability-induced dynamics of multi-component  solitary waves resembles the equation of motion for an effective classical  particle subjected to a $N$-dimensional potential field,
\end{itemize}

\begin{equation} \label{particle}
  E 
  = \frac{1}{2} \sum_{n=1}^N \sum_{m=1}^N 
                   M_{nm} \frac{d \nu_n}{d z} \frac{d \nu_m}{d z} 
  + W(\mbox{\boldmath $\beta$},\mbox{\boldmath $\nu$}).
\end{equation}
Here $M_{nm}$ are the elements of the positive-definite ``mass matrix'', 
\[
M_{nm} = \int_{-\infty}^{\infty} d{\bf x} \sum_{k=1}^N G_{kn}({\bf x}) G_{km}({\bf x}),
\]
\[
G_{kn}({\bf x}) = \frac{1}{\Phi_k({\bf x})} \left( \int_0^{\bf x} d {\bf x}' \Phi_k({\bf x}') \frac{\partial \Phi_k({\bf x}')}{\partial \beta_n} \right), 
\]
$\mbox{\boldmath $\nu$}$ is the vector describing a perturbation  to the soliton parameters $\mbox{\boldmath $\beta$}$, and $W(\mbox{\boldmath $\beta$},\mbox{\boldmath $\nu$})$ is an effective potential energy defined as
\[
   W(\mbox{\boldmath $\beta$},\mbox{\boldmath $\nu$}) 
   = \Delta H({\bf \beta}) 
   + \sum_{n=1}^N \left( \beta_n + \nu_n \right)  \Delta P_n({\bf \beta}),
\]
where $\Delta H({\bf \beta}) = H_s(\mbox{\boldmath $\beta$} + \mbox{\boldmath $\nu$}) - H_s(\mbox{\boldmath $\beta$})$,
and $\Delta P_n({\bf \beta}) =  P_{sn}(\mbox{\boldmath
$\beta$} + \mbox{\boldmath $\nu$}) - P_{sn}(\mbox{\boldmath $\beta$})$, where index ``s'' marks the values calculated on the energy surface.
A proof of those results and their comparison with the theorems of Grillakis {\em et al.}~\cite{Gr,Gr1} are presented in Ref.~\cite{rigorous}.

\subsection{ Example: two coupled NLS equations}

In order to demonstrate how this general theory can be applied to a particular physical problem, we consider  the case of {\em two incoherently coupled NLS equations},
\begin{eqnarray} \nonumber
  i \frac{\partial \psi_1}{\partial z} 
  + \frac{\partial^2 \psi_1}{\partial x^2} 
  + \left( |\psi_1|^2 + \gamma |\psi_2|^2 \right) \psi_1 
  & = & 0,  \\                                          \label{couple}
  i \frac{\partial \psi_2}{\partial z} 
  + \frac{\partial^2 \psi_2}{\partial x^2} 
  + \left( |\psi_2|^2 + \gamma |\psi_1|^2 \right) \psi_2 
  & = & 0,
\end{eqnarray}
where $\gamma$ is a coupling parameter. System~(\ref{couple}) is a two-component reduction of the general $N-$component system~(\ref{NLS}) for $d_1 = d_2 = 1$, $\gamma_{11} = \gamma_{22} = 1$, and $\gamma_{12} = \gamma_{21} = \gamma$. An explicit soliton solution of Eq.~(\ref{couple}) can be easily found for $\beta_1 = \beta_2 = \beta$ and $\gamma > -1$ in the form,
\begin{equation} \label{equalamplit}
  \Phi_1(x) 
  = \Phi_2(x) 
  = \sqrt{\frac{2 \beta}{1 + \gamma}} 
    \; {\rm sech} \left( \sqrt{\beta} x \right) .
\end{equation}
This solution describes a two-component solitary wave with the components of equal amplitudes, and it corresponds to a straight line $\beta_1 = \beta_2$ in the parameter plane ($\beta_1,\beta_2$) of a general {\em two-parameter family of solitary waves} of the model~(\ref{couple}). For  $-1 < \gamma \leq 0$,  two-parameter solitons exist everywhere in the plane ($\beta_1,\beta_2$), while for $\gamma > 0$, the soliton existence region is restricted by two bifurcation curves, $\beta_2 = \omega_{\pm}(\gamma) \beta_1$, where
\begin{equation} \label{domain}
  \omega_{\pm}(\gamma) 
  = \left( \frac{\sqrt{1 + 8 \gamma} - 1}{2} \right)^{\pm 2}.
\end{equation}

Approximate analytical expressions for a two-component solitary wave can also be obtained in the vicinity of one of the bifurcation curves~(\ref{domain}), when one of the components of a composite solitary wave becomes small, while the other one is described by a scalar NLS equation. From the physical point of view, this corresponds to a situation when one of the component creates an effective waveguide that guides the other component, and it is known to describe the so-called {\em pulse shepherding effect in optical fibers} where the large-amplitude  component plays a role of a shepherding pulse~\cite{bpw}. The composite soliton, that describes, for example, a large pulse $\psi_1$ guiding a small pulse $\psi_2$, can be found in the form~\cite{rigorous},
\begin{equation} \label{decoupled}
  \Phi_1 = R_0(x) + \epsilon^2 R_2(x) + {\rm O}(\epsilon^4), \;\;\;\;
  \Phi_2 = \epsilon S_1(x) + {\rm O}(\epsilon^3),
\end{equation}
and this solution is valid in the vicinity of the bifurcation curve,
\begin{equation} \label{bif}
  \beta_2 
  = \omega_+(\gamma) \beta_1 
  + \epsilon^2 \omega_{2+}(\gamma) \beta_1 
  + {\rm O}(\epsilon^4).
\end{equation}
The main terms of the asymptotic series~(\ref{decoupled}),(\ref{bif}) are defined as
$$
R_0 = \sqrt{2 \beta_1} {\rm sech} \left(\sqrt{\beta_1} x \right), \;\;\;\;
S_1 = \sqrt{\beta_1} {\rm sech}^{\sqrt{\omega_+}}
\left( \sqrt{\beta_1} x \right),
$$
and
$$
\omega_{2+} = \frac{ \int_{-\infty}^{\infty} dx \left( S_1^4 +
2 \gamma R_0 R_2 S_1^2 \right)}{\int_{-\infty}^{\infty} dx S_1^2},
$$
and the second-order correction $R_2(x)$ is a solution
of the equation,
$$
\left[ - \partial_x^2 + \beta_1 - 6 \beta_1 \; {\rm sech}^2 \left(
\sqrt{\beta_1} x \right) \right] R_2 = \gamma R_0 S_1^2.
$$

From the existence domain of this two-component soliton,  it follows
that $\omega_{2+}(\gamma) > 0$,  for $0 < \gamma < 1$,  and
$\omega_{2+}(\gamma) < 0$,  for $\gamma > 1$. At $\gamma = 1$
(the so-called {\em Manakov model}), a family of two-parameter composite
solitons becomes degenerated: it exists on the line $\beta_1 = \beta_2$
but,  generally, it is different from the one-parameter solution~(\ref{equalamplit}).  The coupled solitons are known to be stable for the integrable case  $\gamma = 1$. Here we apply the stability theory developed above  and prove that the (1+1)-dimensional two-parameter solitons, including the solitons of equal amplitudes~(\ref{equalamplit}), are {\em stable} for $\gamma \geq 0$, and {\em unstable} for $\gamma < 0$.

Following Ref.~\cite{rigorous} and Sec.~\ref{sec:multi_comp_general}, we evaluate the indices $p(U)$ and $n(\Lambda)$ for the explicit solution~(\ref{equalamplit}). As follows from Eqs.~(\ref{couple}) and~(\ref{equalamplit}),  the Hessian matrix ${\bf U}$ with
the elements~(\ref{Hessian}) can be found in the form,
$$
\frac{\partial P_1}{\partial \beta_1} =
\frac{\partial P_2}{\partial \beta_2}
= \frac{1}{\sqrt{\beta} ( 1 + \gamma)} \;\;\;\mbox{and}\;\;\;
\frac{\partial P_1}{\partial \beta_2} =
\frac{\partial P_2}{\partial \beta_1}
= -\frac{\gamma}{\sqrt{\beta} ( 1 + \gamma)}.
$$
Therefore, the Hessian matrix has $p(U) = 2$
positive eigenvalues,  for $-1 < \gamma < 1$, and $p(U) = 1$ positive
eigenvalue,  for $\gamma > 1$. On the other hand, the linear matrix
operator ${\bf L}_1$ given below Eq.~(\ref{second}) can be diagonalized for
linear combinations of the eigenfunctions,  $v_1 = u_1 + u_2$ and $v_2 =
u_1 - u_2$, such that
\begin{eqnarray}
\nonumber
\left[ - \partial^2_x + \beta - 6 \beta \; {\rm sech}^2
\left(\sqrt{\beta} x \right) \right] v_1 & = & \mu v_1, \\
\label{diagonal}
\left[ - \partial^2_x + \beta - 2 \beta
\frac{(3 - \gamma)}{(1 + \gamma)} \;
{\rm sech}^2 \left(\sqrt{\beta} x \right) \right] v_2 & = & \mu v_2.
\end{eqnarray}
Both the operators in Eqs.~(\ref{diagonal}) are the linear Schr\"{o}dinger operators with solvable sech-type
potentials, and the corresponding eigenvalue spectra are
well studied. The first operator always has a single
negative eigenvalue for $\mu = -3 \beta$, whereas the second operator  has no negative eigenvalues, for $\gamma > 1$, has a single negative eigenvalue, for $0 < \gamma < 1$, and it has
two negative eigenvalues, for $-1 < \gamma < 0$. Thus, in total there
exist $n(\Lambda) = 3$ negative eigenvalues,  for
$-1 < \gamma < 0$, $n(\Lambda) = 2$ negative eigenvalues,
for $0 < \gamma < 1$, and $n(\Lambda) = 1$ negative eigenvalue, for $\gamma > 1$.

Applying the stability and instability results of Ref.~\cite{rigorous} summarized in Sec.~\ref{sec:multi_comp_general}, we come to the conclusion that  the soliton solution~(\ref{equalamplit}) with equal amplitudes is {\em linearly stable} for $\gamma > 0$, since in this domain $p(U) = n(\Lambda) = \{1,2\}$, and it is {\em linearly unstable} for $-1 < \gamma < 0$, since in this domain $p(U) = 2 < n(\Lambda) = 3$.

Finally, we mention the other limiting case that describes the shepherding effect [see  Eqs.~(\ref{decoupled}) and~(\ref{bif})]. In this case, the elements~(\ref{Hessian}) of the  Hessian matrix
${\bf U}$ can also be calculated in an explicit analytical form, and it can be shown that the Hessian matrix ${\bf U}$ has $p(U) = 2$ positive eigenvalues, for $0 < \gamma < 1$, and $p(U) = 1$ positive eigenvalue, for $\gamma > 1$.

On the other hand, the linear matrix operator ${\bf L}_1$ cannot be diagonalized for the soliton in the shepherding regime~(\ref{decoupled}), unless $\epsilon = 0$. In the latter, decoupled, case, it has a single negative eigenvalue at $\mu = - 3 \beta_1$ and a double degenerate zero eigenvalue. When $\epsilon \neq 0$, the zero eigenvalue shifts to become $\mu = - 2 \omega_{2+}(\gamma) \beta_1 \epsilon^2 + {\rm O}(\epsilon^4)$.
Therefore, the matrix operator ${\bf L}_1$ for the soliton~(\ref{decoupled}) has $n(\Lambda) = 2$ negative eigenvalues, for $0 < \gamma < 1$, and $n(\Lambda) = 1$ negative eigenvalue,  for $\gamma > 1$. Thus, we come to the conclusion that {\em the soliton in the shepherding regime is stable} for $\gamma > 0$ since $p(U) = n(\Lambda) = \{1,2\}$.

\section{Multi-hump solitary waves} \label{sec:mhump}

In the simplest case, a spatial optical soliton is  created by one beam of a certain polarization and frequency, and it can  be viewed as a self-trapped mode of an effective waveguide it induces in a  bulk medium~\cite{chiao}.  When a spatial soliton is composed of two (or  more) modes of the induced waveguide~\cite{snyder_dyn}, its  structure becomes rather complicated, and the soliton  intensity profile may  display several peaks. This happens when one of the components creates an effective waveguide that guides higher-order modes of the other component. The resulting solitary waves are  usually referred to as {\em  multi-hump solitary waves}. Their modal structure is more complex than that of multi-component fundamental solitons considered in Sec.~\ref{sec:multi_comp}.
 
For a long time, it was believed that {\em all types} of
multi-hump  solitary waves {\em are linearly unstable}, except for the special case of neutrally stable solitons in the integrable Manakov model. On the contrary, the experimental work of Mitchell {\em et al.}~\cite{prl_multi} reported the observation of stationary structures resembling multi-hump solitary waves. Moreover, the recent analysis~\cite{multi} confirmed that multi-hump solitons supported by incoherent interaction of two optical beams in a photorefractive medium can indeed be stable. Below, we address this issue following the original results by Ostrovskaya {\em et al.}~\cite{multi}.    

In the experiments~\cite{prl_multi}, spatial multi-hump solitary waves were generated  by incoherent interaction of two optical beams in a biased photorefractive  medium. The corresponding model is described by a system of  two coupled nonlinear equations for the normalized beam envelopes, $u(x,z)$ and $w(x,z)$, 
\begin{equation} \label{eq_uw}
  \begin{array}{l} 
  {\displaystyle
   i \frac{\partial u}{\partial z}
   + \frac{1}{2}\frac{\partial^2  u}{\partial x^2}
   + \frac{ u(| u|^2+| w|^2)}{1 + s(| u|^2+| w|^2)}
   - u
   = 0,
  } \\*[9pt] {\displaystyle
   i \frac{\partial w}{\partial z}
   + \frac{1}{2}\frac{\partial^2  w}{\partial x^2}
   + \frac{ w(| u|^2+| w|^2)}{1 + s(| u|^2+| w|^2)}
   - \lambda w
   = 0,
  } \end{array}
\end{equation}
where the transverse, $x$, and propagation, $z$, coordinates are measured in the units of $(L_d/k)^{1/2}$ and $L_d$, respectively, where $L_d$ is a diffraction length and  $k$ is the wavevector. The parameter $\lambda$ has a meaning of a ratio of the nonlinear propagation constants $\beta_1$ and $\beta_2$ of two components, and $s$ is an effective {\it saturation parameter} (see also Sec.~\ref{sec:symm_break} below).

We look for stationary, $z$-independent, solutions of Eqs.~(\ref{eq_uw}) with both components  $u(x)$ and  $w(x)$ real and vanishing as $|x| \rightarrow \infty$. Different types of such two-component localized
solutions, existing for $0<\{\lambda,s\}<1$, can be characterized by the
total power, $ P(\lambda,s)=P_u+P_w$, where $P_u=\int_{-\infty}^{\infty}|u|^2 dx$ and $P_w=\int_{-\infty}^{\infty}|w|^2
dx$. If one of the components is small (e.g., $w/u \sim \varepsilon$),
Eqs.~(\ref{eq_uw}) become decoupled and, in the leading order, the equation
for the $u$-component has a solution $u_0(x)$ in the form of a fundamental, $sech$-like, soliton with no nodes. The second equation can then be
considered as an eigenvalue  problem for the ``modes'' $w_n(x)$ of a waveguide created by the soliton  $u_0(x)$ with the  effective refractive  index profile $u^2_0(x)/[1+s u^2_0(x)]$. Parameter $s$  determines the  total number of guided modes and the cut-off value for each  mode,  $\lambda_n(s)$. Therefore, a two-component vector  soliton $(u_0,w_n)$ consists of a fundamental soliton and a $n$-order mode of  the waveguide it induces in the medium, and it can be characterized by  its ``state vector'': $|0,n \rangle$.

\begin{figure}
\includegraphics[width=.8\textwidth]{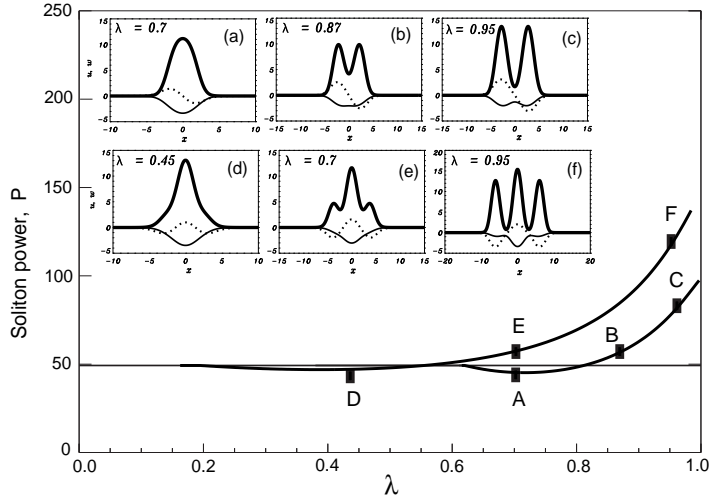}
\caption{ \label{fig:mhump1}
Soliton bifurcation diagram at  $s=0.8$.  Horizontal line -
fundamental $u-$soliton. Curve A-B-C~--- branch of  $| 0,1\rangle$ solitons. Curve D-E-F~--- branch of $| 0,2\rangle$ solitons. Inset: Transverse profiles of $u$ (thin) and $w$ (dashed) fields, and total intensity (thick), shown for marked points~\cite{multi}}
\end{figure}

On the $P(\lambda)$ diagram, continuous branches representing $|0,n \rangle$ solitons emerge at the  points of bifurcations $\lambda_n(s)$ of
one-component solitons. Families of two-component composite solitons are found by numerical relaxation technique~\cite{multi}, and some results of these calculations are presented in Fig.~\ref{fig:mhump1}, for $|0,1 \rangle$ and $|0,2\rangle$ solitons at $s=0.8$. Observing the modification of soliton profiles with changing $\lambda$ (see inset in Fig.~\ref{fig:mhump1}), one can see that the modal description of two-component solitons is valid only near the points
of bifurcations. For $\lambda \gg \lambda_n$, the amplitude of an initially  small $w$-component {\em grows} and the soliton-induced waveguide deforms.  Two- and three-hump solitons are members of the soliton families $|0,1 \rangle$ (branch A-B-C) and $|0,2\rangle$ (branch D-E-F) originating at different bifurcation points. At $\lambda \sim \lambda_n(s)$, while the $w$-component remains small, all $| 0,n\rangle$ solitons are {\it single-humped}, as shown in Figs. 6(a,d). As the amplitude of the component $w$ grows with increasing $\lambda$, the total intensity profile, $I(x)= u^2(x)+w^2(x)$, develops $(n+1)$ humps [see Figs. 6(b,e)], and at sufficiently large $\lambda$ the $u$-component itself becomes {\it multi-humped} [Figs. 6(c,f)]. Separation distance between the soliton humps tends to infinity as $\lambda \rightarrow 1$.

To analyze the linear stability of these multi-hump solitons, we seek solutions of  Eqs.~(\ref{eq_uw}) in the form of weakly perturbed solitary waves:
$u(x,z)=u_0(x)+\varepsilon [F_u(x,z)+iG_u(x,z)]$ and $w(x,z)=w_n(x)+\varepsilon [F_w(x,z)+iG_w(x,z)]$, where
$\varepsilon \ll 1$. Setting  $F_{u,w}\sim f_{u,w}(x)e^{\mu z}$, $G_{u,w}\sim g_{u,w}(x)e^{\mu z}$, one can obtain the following eigenvalue problems
\begin{eqnarray}
  \hat{\mathcal{L}}_1 \hat{\mathcal{L}}_0 \vec g 
  = - \Lambda \vec g 
     , \label{evp1} \\
  \hat{\mathcal{L}}_0 \hat{\mathcal{L}}_1 \vec f
  = - \Lambda \vec f
     . \label{evp2} \nonumber
\end{eqnarray}
Here $\vec g \equiv (g_u,g_w)^T$, $\vec f \equiv (f_u,f_w)^T$, $\Lambda = \mu^2$, and
\[
\hat{\mathcal{L}}_{0,1} = \left(\begin{array}{cc}
-\frac{1}{2}\frac{d^2}{d x^2} +1-a_{0,1} & b_{0,1}\\
b_{0,1} & -\frac{1}{2}\frac{d^2}{d x^2} +\lambda-c_{0,1}
\end{array}\right),
\]
where $a_0=c_0=I/(1+sI)$, $b_0=0$, $a_1=a_0+2u_0^2/(1+sI)^2$,
$c_1=c_0+2w_n^2/(1+sI)^2$, and $b_1=-2u_0w_n/(1+sI)^2$.

\begin{figure}
\sidecaption
\includegraphics[width=.5\textwidth]{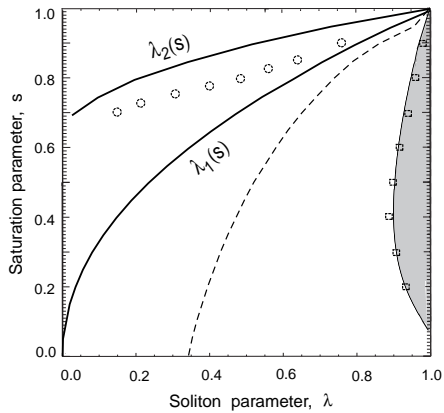}
\caption{ \label{fig:mhump2}
Existence and stability domains for two- and three-hump solitons. Shown are the existence thresholds $\lambda_1(s)$ and $\lambda_2(s)$ for $| 0,1\rangle$ and $| 0,2 \rangle$ soliton families. Dashed~--- the line where total intensity of $|0,1\rangle$ soliton develops humps. Shaded~--- instability domain for two-hump solitons determined by generalized Vakhitov-Kolokolov criterion. Squares and circles~--- instability threshold for two- and three-hump solitons, correspondingly, obtained by numerical solution of the eigenvalue problem~\cite{multi}}
\end{figure}

Note that $\hat{\mathcal{L}}_{1}\hat{\mathcal{L}}_{0}$ and $\hat{\mathcal{L}}_{0}\hat{\mathcal{L}}_{1}$ are adjoint operators with identical spectra. Therefore, we can consider the spectrum  of only one of
this operators, e.g. $\hat{\mathcal{L}}_{1}\hat{\mathcal{L}}_{0}$.
Considering the complex $\Lambda$-plane, it is straightforward to show that
$\Lambda\in (-\infty,-\mu^2)$ is a continuum part of the spectrum with unbounded eigenfunctions. Stable bounded eigenmodes of the discrete spectrum, i.e. the {\em soliton internal modes}, can have eigenvalues only inside the gap, $-\mu^2 < {\rm Re} \, \Lambda <0$. According to the general theory outlined in Sec.~\ref{sec:linear_eigen}, the presence of either positive $\Lambda$ or a pair of complex conjugated $\Lambda$'s implies soliton instability.

Numerical solution of the eigenvalue problem~(\ref{evp1}) shows that both types of solitary wave solutions  {\it can be stable in a certain region of their existence domain}, see Fig.~\ref{fig:mhump2}. In the case of two-hump solitons, the appearance of the instability is related to the fact that close to the curve where the total intensity of the soliton becomes two-humped (see Fig.~\ref{fig:mhump2}, dashed line), a pair of the soliton internal modes split from the continuum spectrum into the gap. 

With the aid of analytical asymptotic technique~\cite{P1}, it is possible to show that a perturbation mode with small but positive  eigenvalue, and therefore linear instability of a general localized solution $(u,w)$, appears if the determinant $J\{u,w\}$ defined as
\[
  J = \frac{\partial (P_u,P_w)}{\partial (\beta_1, \beta_2)} =
  \frac{P_u}{2s}\frac{\partial P_w}{\partial \lambda} 
  - \frac{P_w}{2s}\frac{\partial P_u}{\partial \lambda} 
  + \frac{\partial P_u}{\partial s} \frac{\partial P_w}{\partial \lambda}
  - \frac{\partial P_w}{\partial s} \frac{\partial P_u}{\partial \lambda},
\]
changes its sign. Instability threshold given by the condition $J=0$ is, in fact, the Vakhitov-Kolokolov stability criterion~\cite{VK} generalized for the case of two-parameter composite solitons as described in Sec.~\ref{sec:multi_comp}. However, such a generalization is rigorous for fundamental solitons only, whereas here we deal with higher-order solitary waves. As a result, in this case the determinant condition does not necessarily give a threshold of leading instability and,  therefore, the presence of other instabilities (which are not associated with the bifurcation condition $J=0$ and can even have larger growth rates) is still possible. As was shown above, this is indeed true for the three-hump solitons, whereas the stability condition of two-hump solitons is well defined by the determinant condition $J=0$, as verified by numerical solution of the linear eigenvalue problem (see Fig.~\ref{fig:mhump2}).

\section{Oscillatory instabilities}

\begin{figure}
\sidecaption
\includegraphics[width=.5\textwidth]{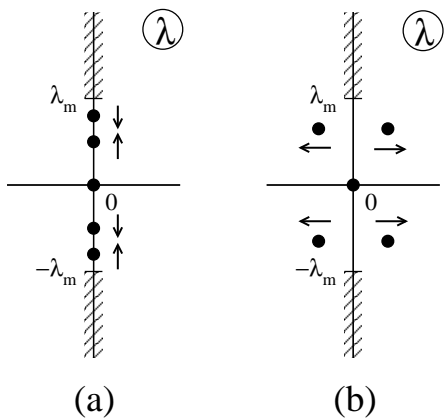}
\caption{ \label{fig:oscillatory_collision}
Schematic presentation of one of the scenarios of the soliton oscillatory instability that occurs via collision of two soliton internal modes}
\end{figure}

When the conditions of the Vakhitov-Kolokolov criterion~\cite{VK} or the theorem of Grillakis {\em et al.}~\cite{Gr,Gr1} and their generalizations discussed in the preceding sections are not satisfied, for example, when there exists more than one negative eigenvalue of the linear eigenvalue problem in a scalar case, the study of soliton stability should be based on the results of numerical simulations, and no simple stability criterion that involves the system invariants can be constructed. In such a case unstable eigenvalues are {\em complex} and, therefore, the corresponding soliton instability is called oscillatory. {\em Oscillatory instabilities} can be usually associated with resonances between two (or more) soliton internal modes, or resonances between the soliton internal mode and the edge of the continuous radiation spectrum. The former case is schematically presented in Fig.~\ref{fig:oscillatory_collision}.

One of the first examples of oscillatory instabilities of solitary waves was found and analyzed by Barashenkov {\em et al.}~\cite{gap} for the case of optical gap solitons described, in the framework of the coupled-mode theory, by two nonlinear coupled equations for the dimensionless amplitudes of the forward- and backward-propagating waves,
\begin{eqnarray} \nonumber
  i \left(\frac{\partial u}{\partial z} 
  + \frac{\partial u}{\partial x} \right) 
  + \left( |u|^2 + \sigma |v|^2 \right) u 
  + v 
  & = & 0,  \\
                                               \label{gap_eq}
  i \left(\frac{\partial v}{\partial z} 
  - \frac{\partial v}{\partial x} \right) 
  + \left( |v|^2 + \sigma |u|^2 \right) v 
  + u 
  & = & 0,  
\end{eqnarray}
where $\sigma$ is the cross-phase-modulation parameter, and the case $\sigma = 0$ corresponds to the exactly integrable massive Thirring model of the field theory.

The fact that the system~(\ref{gap_eq}) is integrable at $\sigma = 0$ can be employed~\cite{gap} to study bifurcations of new eigenvalues from the edge of the continuous spectrum in the case of small but nonzero $\sigma$, similar to the case of a scalar weakly perturbed cubic NLS equation (see Sec.~\ref{sec:imode}).  Additional eigenvalues correspond to internal modes of a gap soliton, and collision of two such eigenvalues results in a soliton oscillatory instability characterized by a pair of complex eigenvalues with a positive real part.

Existence of soliton oscillatory instabilities is usually associated with multi-parameter solitary waves such as optical gap solitons  (in the latter case, two independent parameters are the soliton frequency and velocity).  Another example of the soliton oscillatory instability was recently discussed by Mihalache {\em et al.}~\cite{mih} for the soliton propagation in birefringent optical fibers in the presence of walk-off, self-phase modulation, cross-phase modulation, and parametric four-wave mixing effects.  As was found, the condition of linear marginal stability of the two-parameter solitary waves of that model is not necessarily given by an explicit determinant criterion, because oscillatory instabilities associated with complex eigenvalues were also found to exist in that model.

One more interesting example of oscillatory instabilities of solitary waves was found for {\em dark solitons} in the generalized Ablowitz-Ladik equation and the discrete NLS model of the form~\cite{johan}
\begin{equation} \label{discrete}
  i \frac{\partial \psi_n}{\partial z} 
  + C(\phi_{n+1} + \psi_{n-1}) 
  + |\psi_n|^2 \psi_n = 0.
\end{equation}
It was revealed that even weak inherent discreteness of a nonlinear model can lead to instabilities of solitary waves and, in particular, in the model~(\ref{discrete}) the instability of a dark soliton on a fixed background, $|\psi_{\infty}|^2 =1$, appears for $C > C_{\rm cr} \approx 0.076$, and it is of the oscillatory type. Additionally, it was found that the oscillatory instabilities may appear in a result of {\em two different scenarios}, and they may occur due to either a resonance between radiation modes and the soliton internal mode, or a resonance between two soliton internal modes (as shown in Fig.~\ref{fig:oscillatory_collision}). 

\section{Symmetry-breaking instabilities} \label{sec:symm_break}

All the problems discussed above involve the perturbations acting on a soliton in the space of the same dimension.  However, a solitary wave may also become unstable being subjected to the action of perturbations of higher dimensions. We define these cases as {\em symmetry-breaking soliton instabilities}, which can be further classified into two categories: {\em transverse instabilities} and {\em modulational instabilities}. These two categories are characterized by the type of the soliton evolution affected either by higher spatial dimensions (such as break-up of a plane soliton) or temporal modulation (such as the temporal instability of spatial solitons and the formation of {\em light bullets}).

A comprehensive overview of the soliton self-focusing and  transverse instabilities has been recently presented by Kivshar and Pelinovsky~\cite{phys-rep}, and we refer to that review paper for detailed discussions of a general theory of the symmetry-breaking (both transverse and modulational) soliton instabilities and different examples of the instability-induced soliton dynamics, in the case when a plane solitary wave is subjected to perturbations of higher dimensions. Less studied case is the so-called {\em azimuthal instability} of radially symmetric solitary waves, recently observed experimentally for a number of physical systems. Below we present one of the examples of such symmetry-breaking instabilities.

First of all, we recall that in higher dimensions there exist different types of radially symmetric ring-like solitary waves of higher order, even in a scalar NLS equation with generalized nonlinearity~\cite{higher}. Such modes consist of a bright central spot surrounded by one or more rings, and all those structures are known to be unstable evolving into a fundamental radially symmetric soliton and radiation, or displaying more complicated decay-induced collapse-free evolution in a saturable nonlinear medium~\cite{edm}.

More interesting types of radially symmetric localized modes are associated with a nonzero angular momentum of the solitary waves carrying a topological charge (or ``spin'') $m$, $u(r, \phi) = u(r)  e^{im \phi}$, where $m = \pm 1, \pm 2, \ldots$. Such solitary waves are known as {\em vortex solitons} (see, e.g., Ref.~\cite{kruglov}), and they were found for different types of nonlinear models (see, eg. Ref.~\cite{vortex-gen}). All types of these radially symmetric ring-like solitary waves are also {\em unstable} in respect to azimuthally dependent perturbations, and the decay scenario is determined by the value of angular momentum. In particular, due to modulational instability the rings with the charge $m = \pm 1$ decay into pairs of fundamental bright solitons (``filaments'') of the opposite phase, that strongly repel each other and always move apart along straight trajectories satisfying the momentum conservation~\cite{vortex-gen}. Different filamentation scenarios were observed in experiments on solitons in Rb vapors~\cite{exp}. 

In the scalar beam propagation, symmetry-breaking instabilities of higher-order radially symmetric solitary waves (with or without an angular momentum) usually lead to a soliton decay or filamentation. However, for the vectorial (or multi-component) nonlinear modes, the decay scenarios of ring-like solitary waves become more complicated and lead to novel phenomena. One of the recent examples of this kind is the prediction of the existence and stability of a novel type of robust vector soliton~--- {\em a dipole-mode vector soliton}, a radially {\em asymmetric} two-component self-trapped beam in a bulk medium that can be created via the symmetry-breaking instability of a vortex-mode composite soliton earlier predicted in Ref.~\cite{vortex}.

To discuss this phenomenon, we follow the original work~\cite{dipole} and consider two incoherently interacting beams that propagate along the direction $z$ in a bulk saturable nonlinear medium. This model corresponds, in the so-called isotropic approximation, to the experimentally realized self-trapped beams in photorefractive crystals, and in the dimensionless form it can be described by a system of two coupled NLS equations for the slowly varying beam envelopes $E_1$ and $E_2$, 
\begin{equation} \label{sNLS}
  i \frac{\partial E_{1,2}}{\partial z} 
  + \Delta_{\perp} E_{1,2} 
  - \frac{E_{1,2}}{1 + |E_1|^2 + |E_2|^2} 
  = 0,
\end{equation}
where $\Delta_{\perp}$ is the transverse Laplacian. Stationary solutions of Eq.~(\ref{sNLS}) can be found in the form:
\[ 
  E_1 = \sqrt{\beta_1} u(x,y) e^{i\beta_1z}, \;\; 
  E_2 = \sqrt{\beta_1} v(x,y) e^{i\beta_2 z},
\]
where $\beta_1$ and $\beta_2$ are two independent propagation constants. Measuring the transverse coordinates in the units of $\sqrt{\beta_1}$, and introducing the ratio of the propagation constants, $\lambda = (1-\beta_1)/(1-\beta_2)$, from Eq.~(\ref{sNLS}) we obtain the normalized stationary equations for $u$ and $v$,

\begin{equation} \label{sys}
 \begin{array}{l}
 {\displaystyle
   \Delta_{\perp} u 
   - u 
   + u f(I) 
   = 0 ,  
 }     \\*[9pt] {\displaystyle
   \Delta_{\perp} v 
   - \lambda v 
   + v f(I) 
   = 0 , 
 } \end{array}
\end{equation}
where $f(I) = I(1+sI)^{-1}$, $I= u^2 + v^2$, and the parameter $s = 1-\beta_1$ plays the role of an effective saturation parameter, as in Eq.~(\ref{eq_uw}), Sec.~\ref{sec:mhump}.

\begin{figure}
\includegraphics[width=.8\textwidth]{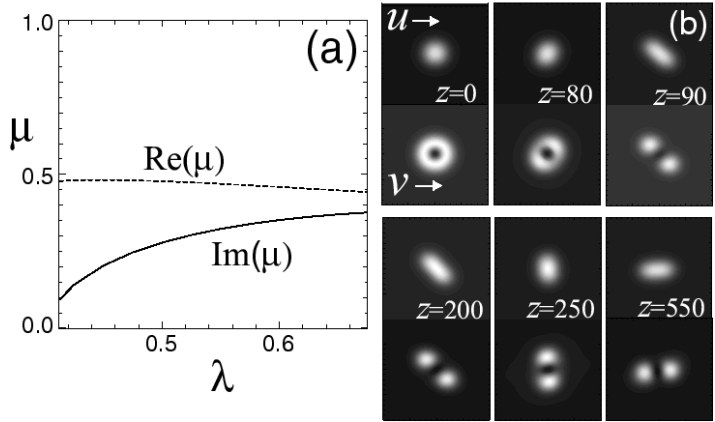}
\caption{ \label{fig:dipole}
(a) Eigenvalue of the leading unstable mode for the vortex-mode vector soliton ($s =0.5$),  and (b) typical decay scenario of the vortex-mode soliton near cutoff ($s=0.65$, $\lambda =0.6$). Shown are intensity distributions of the $u$ and $v$ components in the $(x,y)$ plane~\cite{dipole}}
\end{figure}

Radially symmetric vortex-mode vector solitons of the model~(\ref{sys}) can be found in the form~\cite{vortex}:
\[
  u(x,y) = u(r), \;\;\; v(x,y) = v(r) e^{im\phi},
\]
where $m$ is an integer topological charge. From the physical point of view, such self-trapped states can be regarded as composite solitons consisting of a self-induced waveguide created in the field $u$, and the first-order doughnut-type guided mode in the field $v$, the latter resembling the structure of a Laguerre-Gaussian (LG$_0^1$) mode of a cylindrically symmetric optical waveguide.  Importantly, when such a vortex-mode composite soliton propagates in a nonlinear medium, it undergoes a symmetry-breaking oscillatory-type instability, as shown in Figs.~\ref{fig:dipole}(a,b). However, in a sharp contrast to a vortex decay scenario in a scalar NLS model, two fundamental solitons created in result of such a break-up do not move apart but, instead, they remain trapped in a rotating dipole-like structure, as clearly seen in Fig.~\ref{fig:dipole}(b).  This dipole structure corresponds to an asymmetric vector solitary wave of a different type, that can be thought of as a composite self-trapped state of a soliton-induced waveguide that guides a dipole mode with a shape of a Hermite-Gaussian (HG$_{10}$) laser mode. Recent studies revealed that such dipole-mode solitary waves are indeed extremely robust objects that survive moderate and large perturbations~\cite{dipole}.  The dipole-mode composite solitons have been recently created experimentally in strontium barium niobate (SBN) photorefractive crystals both by a phase-imprinting technique, and as a product of the symmetry-breaking instability of the radially symmetric vortex-mode solitons~\cite{exper,dipole_exper2}.

More recently, the unique robustness of the dipole-mode soliton was demonstrated, theoretically and experimentally,  in the study of their collisions with scalar solitons and other dipoles~\cite{collision}.

\section{Concluding remarks}

We have presented a brief overview of the key concepts and basic results of the soliton stability theory applied to spatial optical solitons, emphasizing the most recent results on the stability of multi-component solitary waves, the study of oscillatory instabilities of solitary waves, and symmetry-breaking instabilities of solitons in higher dimensions.   We have demonstrated that the theory of soliton instabilities is closely related to the concept of soliton internal modes, an important feature of solitary waves of many nonintegrable nonlinear models. 

The simplest and most familiar class of the soliton instabilities in the NLS-type nonlinear models is described by the Vakhitov-Kolokolov criterion, $dP/d\beta < 0$, where $P$ is the soliton power and $\beta$ is its propagation constant. This criterion is valid for one-component fundamental solitary waves provided the $L_1$ operator of the linear eigenvalue problem possesses exactly one negative eigenvalue. We have shown that a nontrivial generalization of this criterion to the case of multi-component solitary waves is also possible, but it requires more restrictive conditions. 

When the rigorous results of the linear stability theory are not available, a simple asymptotic theory allows to find the condition for the marginal stability point (or, in general, a marginal stability surface) that may be further verified numerically. One of such cases, when the result of an asymptotic theory has no rigorous proof yet, even though being verified numerically, is the stability of two-hump solitary waves created by incoherent interaction of two optical beams in a photorefractive nonlinear medium. The multi-scale asymptotic analysis is valid for the region of the marginal stability, and it is a powerful tool that also allows to study weakly nonlinear effects in the instability-induced soliton evolution and to describe analytically different scenarios of the dynamics of unstable solitons.  

In many other cases, the eigenvalues of the corresponding linear problem are complex, and the study of the soliton stability should be carried out by the numerical analysis of the corresponding linear eigenvalue problem. This is the case of the so-called oscillatory instabilities of solitary waves, for instance those known to exist for gap solitons described by a coupled-mode theory. Such instabilities are not predicted by simple integral criteria,  and they appear due to a resonance between two soliton internal modes or a soliton internal mode and the edge of the continuous spectrum.

Many interesting problems of the soliton stability theory 
and the dynamics of unstable optical solitons can be found in higher dimensions,  where the symmetry-breaking instability of plane or radially symmetric solitary waves brings new features into the soliton dynamics. In particular, we have demonstrated that a vortex-like solitonic mode, that always breaks up into fundamental bright solitons in a scalar model, may evolve, in a vector model, into a rotating dipole-like structure associated with a robust dipole-mode vector soliton.  

The study of instabilities is an important direction in the research of spatial optical solitons that links the theoretical concepts of the beam self-trapping and stationary self-guided beam propagation with the possibility of experimental verifications. After all, wave instabilities are probably the most remarkable physical phenomena that may occur in Nature, and their study allows researchers to reveal many important and fascinating properties of nonlinear physical systems. 

\section*{Acknowledgements}

Many of our colleagues contributed to the topics briefly discussed in this Chapter. We are especially indebted to Dmitry Pelinovsky and Elena Ostrovskaya for fruitful collaboration and their valuable contribution to the results presented here and to our understanding of the soliton stability.

\clearpage
\addcontentsline{toc}{section}{Index}
\flushbottom
\printindex
\end{document}